\documentclass[aps,prl,twocolumn,groupedaddress]{revtex4}

\usepackage{graphicx}% Include figure files
\usepackage{bm}% bold math
\usepackage{graphicx}
\usepackage{subfigure}
\usepackage{latexsym}
\usepackage{placeins}
\usepackage{amstext,amssymb,amsfonts}
\usepackage{epsfig}
\usepackage{color}
\usepackage{titlesec}
\usepackage{amsmath}
\usepackage{mathtools}
\usepackage{float}
\usepackage{nameref}
\usepackage{textcase}
\usepackage{mismath}

\let\originalparagraph\paragraph
\renewcommand{\paragraph}[2][.]{\originalparagraph{#2#1}}

\begin{document}

\title{Avalanche-size distribution of Cayley tree}
\author{Amikam Patron\footnote{patron@g.jct.ac.il}}

\affiliation{Department of Mathematics, Jerusalem College of Technology, Jerusalem 91160, Israel}

\date{\today}

\begin{abstract}
Attacks on networks is a very important issue in developing strategies of eradicating spreads of malicious phenomena in networks, such as epidemics and fake information, which are generally named in the research \textit{networks immunization}. The traditional approach of evaluating the effectiveness of attacks on networks, is focused on measuring some macro parameters related to an entire attack, such as the critical probability of a percolation occurrence in the network $p_c$ and the relative size of the largest component in the network -- the \textit{giant component}, but not considering the attack on a micro perspective, which is the analysis of the node removals, during an attack, themselves, their characteristics and results. In this paper we present and apply the last method of focusing on the micro scale of an attack. Based on the theory of percolation in networks, we analyze the phenomenon of an \textit{avalanche} results due to a single node removal from a network, that is a state in which a removal of one node belongs to the network's giant component causes other nodes to be also disconnected from it, which is a much greater contribution to the fragmentation (immunization) of the network than only the single original node removal itself. Specifically, we focus ourselves on the parameter of the \textit{size} of an avalanche, which is the number of nodes that are disconnected from the giant component due to a single node removal. Relating to a random attack on a network of the type of \textit{Cayley tree}, we derive analytically the distribution of the sizes of avalanches occur during the entire attack on it, until the network is dismantled (immunized) and the attack is terminated.
\end{abstract}

\maketitle

\section{Introduction}
An attack on networks, which is a scenario where a combination of nodes (or links) of a network are removed and become non-functional causing the entire network to be dismantled and non-functional, is an issue that has been studied widely
\cite{cohen-2010,albert-nature-2000,shargel-prl-2003,kim-prl-2003,motter-pre-2002,holme-pre-2002,albert-pre-2004,holme-pre-2007,Schneider-proc-nat-acad-sci-2011,hermann-jrnl-stast-mech-2011,lin-epl-2018,patron-njp-2020,molloy-rsalg-1995,cohen-prl-2000,erdds1959random,erd6s1960evolution,Redner-eurp-phs-j-1998,albert-nature-1999,Barabasi-science-1999,Newman-pnas-2001}. There are two types of attacks on networks, on them the research has been focused: \textit{random attack} and \textit{targeted attack} \cite[Chapter~10]{cohen-2010}. In a random attack, the attacker has no information about the network and its topology. Therefore, the choice of the nodes to be attacked is implemented randomly, with no priority of each node to be attacked before or after another node. In each attack stage, the attacker chooses randomly a node from the set of nodes that were not attacked until that stage, and attacks that node. On the other hand, in targeted attack the attacker has some information about the network and its characteristics, and according to it there are  some priorities of attacking some nodes before other nodes. Therefore, in each attack stage the attacker attacks the node with the highest priority among the set of nodes were not attacked until then. An example of a widely studied targeted attack, is where the attacker has the information of the degree of each of the network's nodes, so in each attack stage the node with the highest degree, among the nodes were not attacked until that stage, is chosen to be attacked.     

A very important issue related to the issue of attacks on networks, is the \textit{robustness} of a network, that is the capability of a network to continue being functional during an attack against it. There are many studies about the way of parameterizing the robustness of networks, some of them are based on comparisons of network's parameters before and after the attack such as a change in the \textit{network diameter} \cite{albert-nature-2000,shargel-prl-2003,kim-prl-2003}, the relative size of the large component of the network \cite{motter-pre-2002}, a change in the \textit{betweeness centrality} \cite{holme-pre-2002}, \textit{connectivity loss} \cite{albert-pre-2004}, and \textit{f-robustness} \cite{holme-pre-2007}, and some of them are based on measuring the mean functionality level of a network during the entire attack like the integral of the network's largest component with respect to time during the attack \cite{Schneider-proc-nat-acad-sci-2011,hermann-jrnl-stast-mech-2011,lin-epl-2018,patron-njp-2020}.

An important method of measuring the robustness of a network is based on \textit{percolation theory} \cite{stauffer-1994,grimmett-1999}. The basic description of percolation theory is related to lattices. It presents a state where the sites or the bonds of a lattice are occupied with probability $p$, and unoccupied with probability $1-p$. For each lattice structure (like BCC, FCC, Honeycomb etc.) there is a critical probability denoted $p_c$, such that if $p>p_c$, a cluster of occupied sites that spans the lattice from one side to the other size, named \textit{spanning cluster} or \textit{infinite cluster}, is exists in the lattice, and in opposite if $p<p_c$ there is no spanning cluster in the lattice \cite[Chapter~2]{stauffer-1994}. If a spanning cluster exists in a lattice, it is called that a \textit{percolation} occurs in the lattice. 

The same state could be presented in networks, where the nodes or the edges of a network are occupied with probability $p$ and unoccupied with probability $1-p$. However, for defining a percolation in random networks, where the linking of the set of nodes by edges is performed randomly, we can not apply the lattices criterion of the existence of a spanning cluster, since due to the network structure there is no meaning of spanning the network from size to size. Therefore, another criterion has been stated for percolation in random networks, which is the existence of a large component of occupied nodes in the network whose size scales with the network's size, i.e. $O(N)$, named \textit{giant component}. That is, a typical property of a random network is a critical probability of nodes' occupation $p_c$, such that if $p>p_c$ a giant component exists in the network, and if $p<p_c$ the network is fragmented to small components and a giant component does not exist. As long as $p_c$ get smaller the network is considered more robust, since in order to fragment it a large fraction of $1-p_c$ of the network's nodes has to be removed.

In \cite{molloy-rsalg-1995,cohen-prl-2000} it was shown that the criterion for percolation in random network, i.e. the existence of a giant component in the network, was generated by the configuration model, is $\kappa=\frac{\langle k^2\rangle}{\langle k\rangle}>2$, where $\langle k\rangle$ is the expectation of the node degree and $\langle k^2\rangle$ is the expectation of the square of the node degree. According to this criterion, it was shown that for random networks under random attack
\begin{equation}
	p_c=\frac{1}{\kappa_0-1}, 
	\label{pc_rand_network}
\end{equation}
where $\kappa_0$ is the value of $\kappa$ before the attack on the network begun \cite{cohen-prl-2000}. From this, the value of $p_c$ for two types of random networks ware studied widely, was calculated: (i) For Erd\H{o}s-R\'{e}nyi networks \cite{erdds1959random,erd6s1960evolution}, where the node's degree $k$ follows a Poisson distribution $P(k)=e^{-\lambda}\frac{\lambda^k}{k!}$, it was shown that $p_c=\frac{1}{\lambda}$. (ii) For Scale-Free networks, a topology that was found in many real networks  \cite{Redner-eurp-phs-j-1998,albert-nature-1999,Barabasi-science-1999,Newman-pnas-2001}, where the node's degree $k$ follows a power-law distribution $P(k)\sim k^{-\gamma}$, such that most of the nodes have a very small degree but there is also a small fraction of nodes with a very high degree named \textit{hubs}, it was shown that for $\gamma>3$, $p_c$ has a finite nonzero value, but for $\gamma\leq3$, $p_c$ approaches $0$ as $N$ tends to infinity, that means that although almost of the network's nodes are removed, there still exists a giant component in the network and the network is considered functional.     

Another type of network that has been studied widely including its robustness and aspects of percolation, is the Bethe-Lattice (BL). BL is a network of the type of an infinite tree (no cycles) where all its nodes have the same degree $Z$. Using Eq. $\left(\ref{pc_rand_network}\right)$, it was shown \cite[chapter 10]{cohen-2010} that under random attack, $p_c$ of BL with $Z$ neighbours of each node is 
\begin{equation}
	p_c=\frac{1}{Z-1}.
\end{equation}
This expression was also derived in \cite{braga-SIAMrev-2005} and in  \cite[chapter 2]{bunde-2012}, applying other methods. Another type of network that is similar to BL but not identical, is the Cayley tree (CT), which is a BL but with a finite number of nodes arranged in a finite number of shells. That is, in a CT there is a unique node from it the tree begins. This node is linked to $Z$ neighbours that are the first shell of the tree. Then each node of the first shell is linked to other $Z-1$ neighbours which all of them are the second shell of the tree, then each node of the second shell is linked to other $Z-1$ neighbours which all of them are the third shell of the tree, etc. This process is terminated in the last shell, where each of the nodes of this shell is not branched again to some other $Z-1$ nodes of some next shell. The critical probability $p_c$ of CT is not the same as of BL, since its \textit{finite-size} that has to be considered \cite[chapter 4]{stauffer-1994}. Below in this paper we present a method for calculating the critical probability $p_c$ of CT.

In general, the robustness of a network is a critical characteristics for its survivability under malicious attacks. However, there are typical states in which an attack on a network is a desired state, that is when an extinction of some negative phenomena spread in it is required, such as an epidemic, fake information and some damages. There are indeed many studies deals with the issue of an extinction of negative phenomena in networks \cite{watts-1998-nature,pastor-pre-2002,dezsHo-pre-2002,liu-ncr-2021}, which is named \textit{network immunization}.

However, almost all the studies about attacks on networks from the perspective of percolation theory, deals with the network's properties related to it on a macro scale, such as the percolation threshold $p_c$ and the giant component's size relative to network's size, but not deal with the attack aspects in a micro scale, that is each of the node removals themselves, its characteristics and results. 

In this paper we present this last approach of analyzing an attack on networks from the perspective of a micro scale as mentioned. We focus ourselves on the phenomenon named \textit{avalanche}, that is a state where an attack on a node that is a part of the giant component, causes it to be disconnected from the giant component, and causes also other nodes that were not directly attacked but are dependent on the attacked node, to be also disconnected from the giant component. The total number of nodes are disconnected from the giant component due to an avalanche, is named the \textit{size} of the avalanche. Focusing ourselves on a network of the type of CT, We develop here analytically the expressions of the probability distribution of the various sizes of avalanches occur during a random attack on this network.

\section{The model}
We state the followings definition for the CT structure: The node from it the tree begins is called the `root' of the tree. A `neighbour' of a node is a node that is linked to it. Walking from a node to its neighbour in a direction of moving away from the root, is called `walking down the tree', and in a direction of moving towards the root is called `walking up the tree'. The `sons' of a node are the node's neighbours 
when walking from it down the tree. A `father' of a node is the node's neighbour when walking from it up the tree. The `descendants' of a node are nodes to them there is a path from it, when walking from it down the tree (son of a node is also its descendant). The `ancestors' of a node are nodes to them there is a path from it, when walking from it up the tree (the father of a node is also its ancestor). The set of all the sons of the root is the first layer of the tree and is called `generation 1'. The set of all the sons of nodes belong to generation 1 is the second layer of the tree and is called `generation 2', etc. The number of layers of the tree is denoted $L$. The nodes of the $L$'th layer are the tree dead ends called the `surface' of the tree. The set of these nodes is called `generation $L$'.

An attack on the network is executed by stages, where in each stage one of the network's node is chosen randomly and attacked. Node that was attacked, is not chosen to be attacked again. The attack ends when a sufficient combination of attacked nodes causes the network to be dismantled, i.e. a phase transition occurs and the network's giant component is fragmented. 

During an attack, a node can be in one of three modes named \textit{white}, \textit{black} and \textit{grey}, defined as follows: 
(i) The \textit{wight} mode -- Before an attack on a network began, all the network's nodes are considered functional and each node is defined to be in a \textit{white} mode, that is a functional mode. (ii) The \textit{black} mode -- During an attack, white node that is attacked is switched from a functional to nonfunctional state, causing it to be switched from a white mode to a \textit{black} mode that is an attacked nonfunctional mode. (iii) The \textit{grey} mode -- When a white node is attacked, all its descendants with a wight mode, i.e. the descendants that have not yet been attacked and were functional before the current attack stage, are switched to a nonfunctional state since the attacked node cut their path up the tree to the root. White node that is switched from functional to nonfunctional state in this way, is accordingly switched to be in a \textit{grey} mode, that is not attacked nonfunctional mode. Regarding the \textit{grey} nodes, we also note that since only node that was attacked is not chosen to be attacked again, it is possible that a grey node, although nonfunctional but still has not been attacked, would be chosen to be attacked. In this case, the attacked grey node is switched to a black mode.
According to these definitions of the three modes, it is guaranteed that the ancestors of a white node can be only white and the descendants of a wight node can be white or black, the ancestors of a grey node can be grey or black and the descendants of a grey node can be grey or black, and the ancestors of a black node can be white, grey or black and the descendants of a black node can be grey or black.
Since the root of the tree has no father, it can be in white mode or black mode only.

`Avalanche' as defined above is a combination of nodes are disconnected from the giant component due to one node removal. Accordingly, an `avalanche' in our model is a change of the mode of a set of wight nodes due to an attack on a wight node, such that the wight attacked node is switched to a black mode, and the wight descendants of the attacked node are switched to a grey mode. `Avalanche size' is the number of white nodes that change their mode in an avalanche. `Complete avalanche' is an avalanche where all the descendants of a white attacked node are also white, so the attacked white node is switched to a black mode and all its descendants down the tree until the surface of the tree are switched to a grey mode. To illustrate it, for a CT with $Z$ neighbours, attacking a white node in the $L$'th layer causes a complete avalanche of size $1$, attacking a white node in the $L-1$ layer where its $Z-1$ sons of the $L$'th layer are in white mode, causes a complete avalanche of size $1+\left(Z-1\right)=\frac{\left(Z-1\right)^2-1}{Z-2}=Z$, attacking a white node in the $L-2$ layer where its $Z-1+\left(Z-1\right)^2$ descendants ($Z-1$ of the $L-1$ layer and $\left(Z-1\right)^2$ of the $L$'th layer) are in white mode, causes a complete avalanche of size $1+\left(Z-1\right)+\left(Z-1\right)^2=\frac{\left(Z-1\right)^3-1}{Z-2}$, and etc. In general, attacking a white node in the $L-m+1$ layer $\left(m=1,2,3,\cdots,L\right)$, causes a complete avalanche of size $\frac{\left(Z-1\right)^m-1}{Z-2}$.
`Non-complete avalanche' is an avalanche but not of the complete avalanche type. It occurs when the descendants of a white attacked node are not all of wight mode (some of them were switched to a black or a grey mode before the current attack on their ancestor).
Attacking a grey node, according to the previous formal definition of an avalanche, causes no avalanche (even before it is attacked it was already disconnected from the giant component). Nevertheless, we define this state as a `null avalanche', means that a node was indeed attacked, but this does not contribute to the reduction of the giant component size. Null avalanche size is 0.

\begin{figure*}[htbp]
	\begin{center}
		\begin{tabular}{cc}
			\hspace{-2em} \includegraphics[scale=0.3]{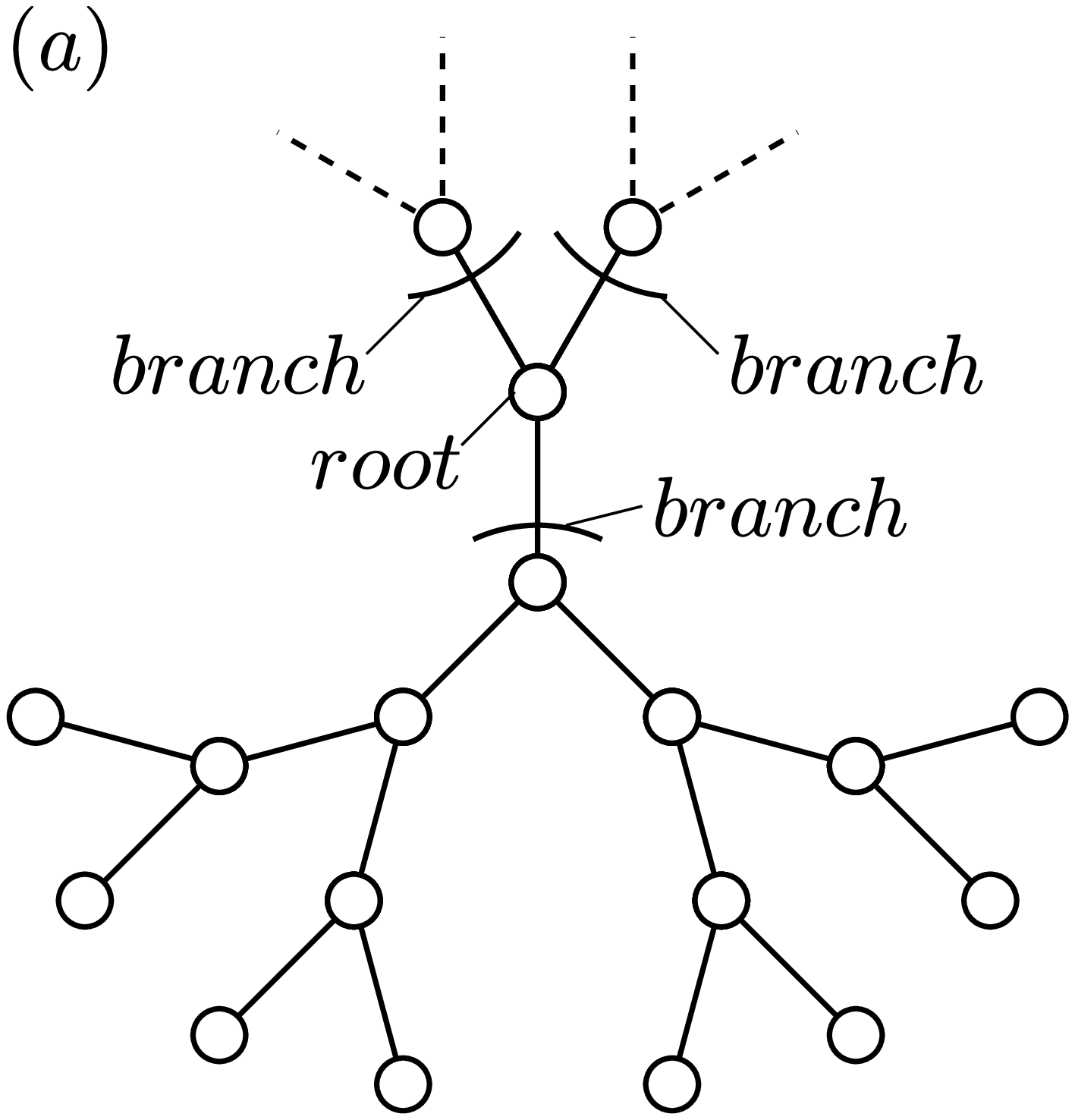}\hspace{-9.2em} &
			\includegraphics[scale=0.3]{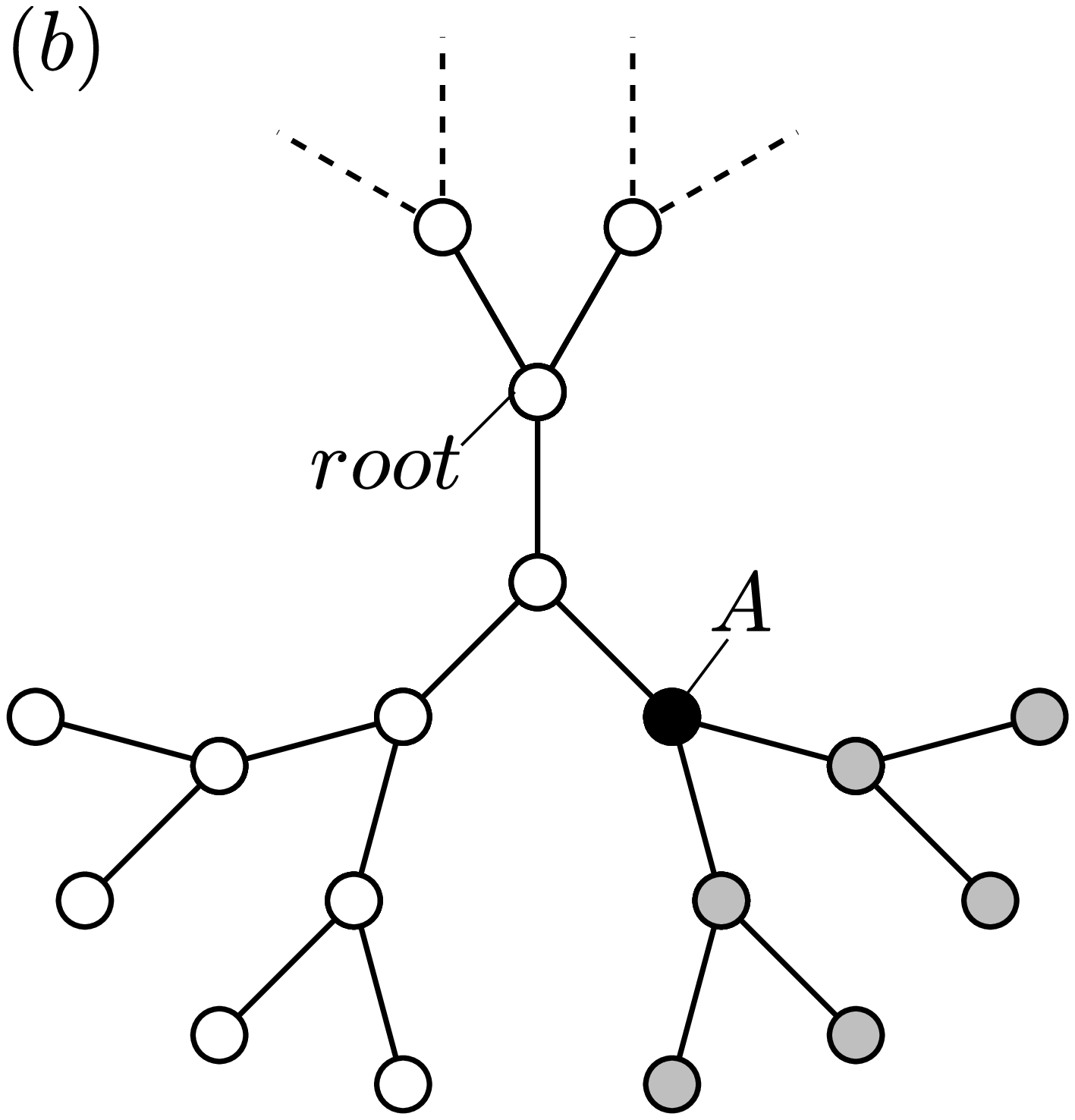} \\
			\hspace{-2em} \includegraphics[scale=0.3]{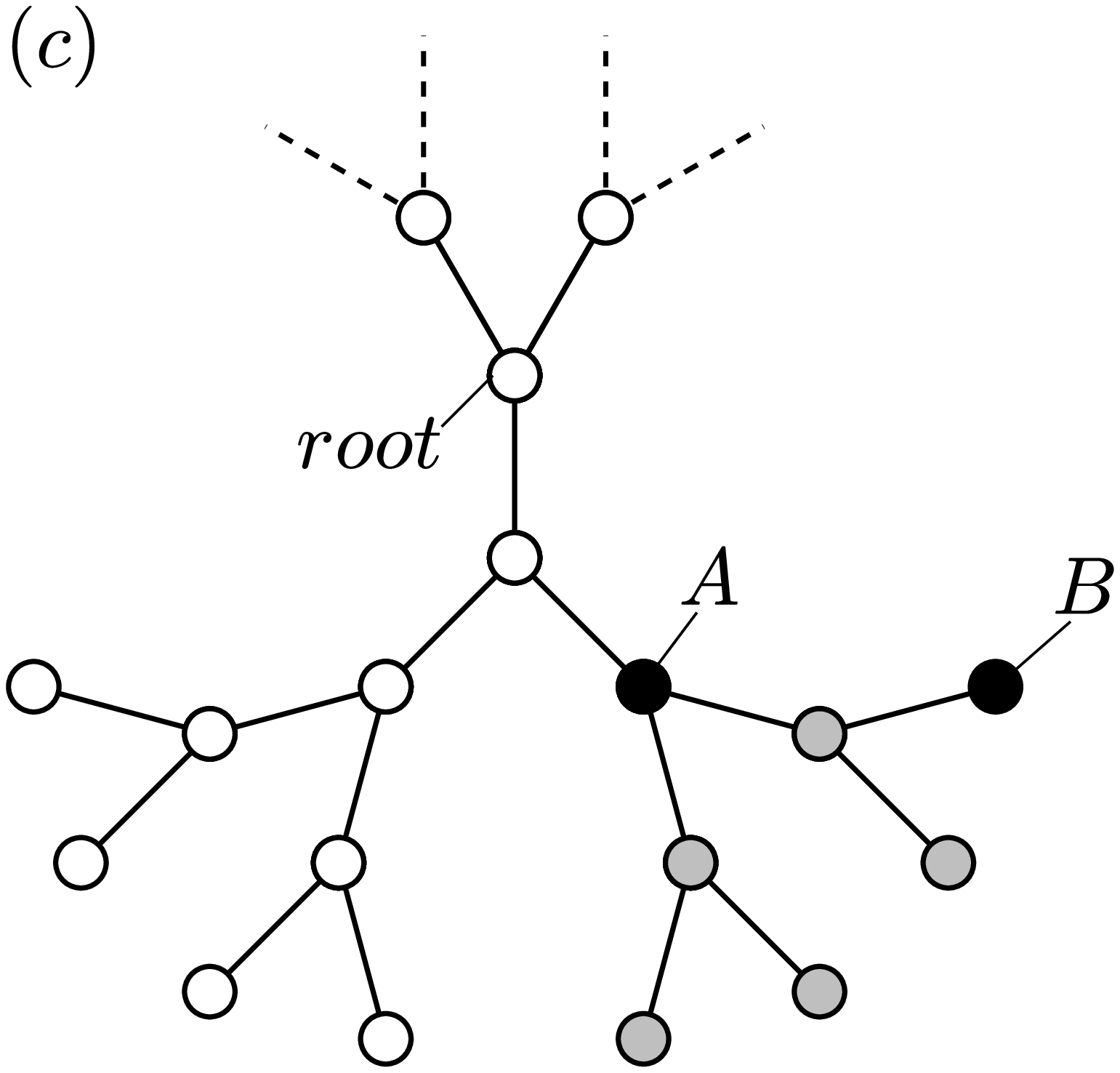}\hspace{-9.2em} &
			\includegraphics[scale=0.3]{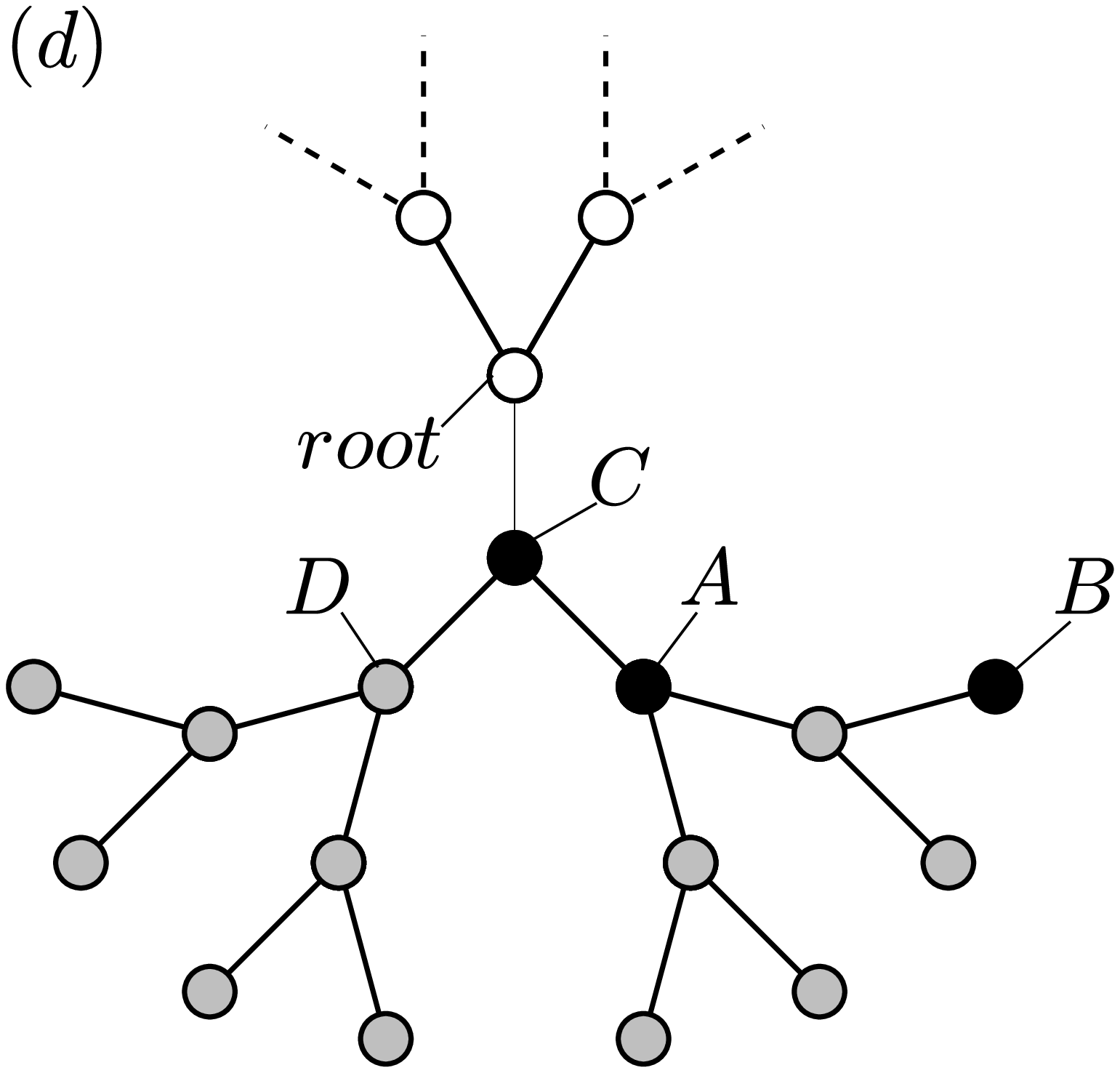} \\
		\end{tabular}
	\end{center}
	\caption{\textbf{Illustration of the model on Cayley tree with $Z=3$ neighbours and $L=3$ layers:} $\left(a\right)$ CT before an attack began. All the nodes are of wight mode. On the sketch are marked the \textit{root} of the tree and the three \textit{branches} emanated from the root. In this figure we focus only on the branch located at the bottom of the sketch, so the nodes belong to this branch are drawn only. $\left(b\right)$ Stage number one of an attack -- The node marked $A$ is attacked. As a result, the node $A$ is switched to a black mode, and its six descendants are switched to a grey mode. Before this stage, all the descendants of node $A$ were of wight mode, so the attack on node $A$ causes a \textit{complete avalanche} of size $7$ (node $A$ and its six descendants). $\left(c\right)$ Stage number two of an attack -- The node marked $B$ is attacked. As a result, it is switched to a black mode. Since before this stage node $B$ was of grey mode, i.e. was disconnected from the root, the attack on node $B$ causes \textit{null avalanche}. $\left(d\right)$ Stage number three of an attack -- The node marked $C$ is attacked. As a result, it is switched to a black mode. Since node $A$ that is a \textit{son} of node $C$ was already attacked before the current stage, then the current stage has no impact on that subbranch of node $C$ contains node $A$ and its descendants. On the other hand, the nodes in the other subbranch of node $C$, which are the node marked $D$ and its descendants, that were of wight mode before the current stage, are all switched to a grey mode. Since before the current stage not all of node $C$ descendants were of wight mode, the current stage causes a \textit{non-complete avalanche} of size $8$ (node $C$, node $D$ and node $D$'s six descendants).}
	\label{fig:model_illustration}
\end{figure*}

Figure \ref{fig:model_illustration} is an illustration of the model. It presents three stages of an attack on CT with $Z=3$ neighbours and $L=3$ layers, including presentations of complete avalanche, non-complete avalanche and null avalanche occur during the attack.

\section{Critical probability $p_c$ of Cayley tree}
\textbf{I. Definitions.} \ We state the following definitions regarding random choices of nodes in the tree: the probability of choosing randomly a black node is denoted by $p_b$. Since in each attack stage only one node is attacked and is switched to a black mode, then between the stages $k$ and $k+1$ of an attack, $p_b=\frac{k}{N}$ where $N$ is the network size.  

The events of choosing randomly a node of white, grey or black mode, are denoted by \textit{W}, \textit{G} and \textit{B}, respectively. The events of choosing a node that is not of white, not of grey or not of black mode are denoted by \textit{$\overline{W}$}, \textit{$\overline{G}$} and \textit{$\overline{B}$}, respectively. The events of choosing randomly a node of the first generation, second generation and so on $\ldots$, until the $L$'th generation are denoted by \textit{$g_1$}, \textit{$g_2$} and so on $\ldots$, until \textit{$g_L$}, respectively.\newline

\textbf{II. Critical probability $p_c$.} \ \ A considered property of CT, resulted due to its structure of a tree with no cycles, is the existence of one path only from any node to the root. Therefore, a percolation in CT is defined as a state where there is at least one path from a node of the tree surface to the root composed of functional nodes, i.e. composed of nodes of wight mode only. As mentioned above, according to our model the descendants of nodes of grey or black mode, could be of grey or black mode only but not of wight mode. Therefore, it guarantees that for each node of wight mode, all of its ancestors until the root are also of wight mode. Specifically, the existence of a node of wight mode belongs to the surface of the tree, guarantees a path from it to the root composed of wight nodes only, i.e. guarantees the fulfillment of the percolation condition. Accordingly, for the sake of calculating $p_c$ of CT, we calculate firstly the probability of choosing randomly a white node given that it belongs to the tree surface. We perform the calculation in some stages as follows: for the first generation -- the event of choosing a node that is not of white mode given that this node is of the first generation, occurs when the father of this node -- the root -- is of black mode, with probability $p_b$, or when the chosen node itself is of black mode and its father -- the root -- is of white mode, with probability $p_b\left(1-p_b\right)$. Thus, the probability of the complement event of choosing a white node given that this node is of the first generation is
\begin{equation}
	P\left(W|g_1\right)=1-\left[p_b+p_b\left(1-p_b\right)\right]=\left(1-p_b\right)^2.
\end{equation} 
In the same way for the second generation -- the event of choosing a node that is not wight given it is of the second generation, occurs when the ancestor of distance 2 of this node -- the root -- is black, with probability $p_b$, or when the father of this node is black and its ancestor of distance $2$ is white, with probability $p_b\left(1-p_b\right)$, or when the chosen node itself is black and its father and its ancestor of distance $2$ are white, with probability $p_b\left(1-p_b\right)^2$. Thus, the probability of the complement event of choosing a node that is wight given that this node is of the second generation is 
\begin{equation}
	P\left(W|g_2\right)=1-\left[p_b+p_b\left(1-p_b\right)+p_b\left(1-p_b\right)^2\right]=\left(1-p_b\right)^3.
\end{equation}
The previous results can be generalized to any generation denoted by $l$, such that the probability of choosing a node that is wight given that this node is of the $l$'th generation is
\begin{equation}
	P\left(W|g_l\right)=\left(1-p_b\right)^{l+1},
	\label{P_W_l}
\end{equation}
and in particular, the probability of choosing white node given it is belonged to the tree surface i.e. belonged to the $L$'th generation is
$\left(1-p_b\right)^{L+1}$.

In a CT with $Z$ neighbours, the number of nodes in the surface is $Z\left(Z-1\right)^{L-1}$. We number the nodes of the surface with the numbers $1,2,3,...,Z\left(Z-1\right)^{L-1}$. We define an indicator random variable $X_j$ for the event of the existence of a path from the node in the surface numbered $j$ to the root, composed of wight nodes only, i.e. the event of percolation due to the $j$'th node of the surface. We define also another random variable $X$ that is the number of paths from the surface to the root composed of wight nodes only, i.e. $X=\sum_{j=1}^{Z\left(Z-1\right)^{L-1}}X_j$. Therefore, the expected number of paths from the surface to the root composed of wight nodes only is
\begin{align}
	E\left[X\right]&=\sum_{j=0}^{Z\left(Z-1\right)^{L-1}}E\left[X_j\right]=\sum_{j=0}^{Z\left(Z-1\right)^{L-1}}\left(1-p_b\right)^{L+1}\nonumber\\
	&=Z\left(Z-1\right)^{L-1}\left(1-p_b\right)^{L+1}.
\end{align}
The condition for percolation is that the expectation of the number of paths from the surface to the root composed of wight nodes only is at least $1$, i.e. $E\left[X\right]\geq1$. The transition occurs when $E\left[X\right]=1$. Accordingly, we define $p_b^*$ as the critical value of $p_b$ at the transition, and get  
\begin{equation}
	Z\left(Z-1\right)^{L-1}\left(1-p_b^*\right)^{L+1}=1.
\end{equation}
Therefore, the critical probability $p_c$ which is the fraction of nodes were not attacked at the critical point of percolation is
\begin{equation}
	p_c=1-p_b^*=\frac{\left(Z-1\right)^\frac{1-L}{1+L}}{Z^\frac{1}{1+L}}.
	\label{p_c_finite}
\end{equation}

Note that when $L>>1$, $p_c$ in Eq. $\left(\ref{p_c_finite}\right)$ tends to be
\begin{equation}
	p_c=\frac{1}{Z-1},
	\label{p_c_approx}
\end{equation}
which is the known condition to percolation of BL. Therefore we conclude that when $L>>1$, the critical probability $p_c$ of CT tends to be the same as of the BL.

In this paper we analyze CT's with a very large $L$, so from now on we consider Eq. $\left(\ref{p_c_approx}\right)$ as the critical probability for percolation in CT taken in consideration.

\section{Avalanche-size distribution of Cayley tree with $Z$ neighbours of each node}
\subsection{Theory}
\subsubsection{Probability of null avalanche}

\textbf{I. Probability of choosing a grey node.} \  \ Recall that null avalanche occurs when a grey node is attacked, and that black nodes that have already been attacked are not chosen to be attacked again, we calculate firstly the probability of choosing randomly a grey node given that the chosen node is not black. 
In order to calculate it, we perform initially a simpler calculation of the probability of choosing randomly a wight node given that the chosen node is not black, in stages as follows: for the first generation -- the event of choosing a node that is not of white mode given that this node is of the first generation and is not of a black mode, occurs when the father of this node -- the root -- is black, and its probability is $p_b$. Thus, the probability of the complement event of choosing a white node given that this node is of the first generation and is not black is
\begin{equation}
	P\left(W|g_1\cap\overline{B}\right)=1-p_b.
\end{equation}  

In the same way for the second generation -- the event of choosing a node that is not wight given it is of the second generation and it is not black, occurs when the ancestor of distance 2 of this node -- the root -- is black, with probability $p_b$, or when the father of this node is black and its ancestor of distance $2$ is white, with probability $p_b\left(1-p_b\right)$. Thus, the probability of the complement event of choosing a node that is wight given that this node is of the second generation and is not black is
\begin{equation}
	P\left(W|g_2\cap\overline{B}\right)=1-\left[p_b+p_b\left(1-p_b\right)\right]=\left(1-p_b\right)^2.
\end{equation}
In the same way the previous results can be generalized for any generation $l$, such that the probability of choosing a node that is wight given that this node is of the $l$'th generation and is not black is
\begin{equation}
	P\left(W|g_l\cap\overline{B}\right)=\left(1-p_b\right)^l.
	\label{P_W_l_nB}
\end{equation}

Since a chosen node has the options of being white or grey only, then the event of choosing a node that is grey given that this node is of the $l$'th generation and is not black, is the complement of the event of choosing a node that is white given that this node is of the $l$'th generation and is not black, and accordingly its probability is
\begin{equation}
	P\left(G|g_l\cap\overline{B}\right)=1-\left(1-p_b\right)^l.
	\label{P_G_l_nb}
\end{equation}
Using the formula of total probability, we get the probability of choosing a node that is grey and is of this $l$'th generation given that this node is not black as follows
\begin{equation}
	P\left(G\cap g_l|\overline{B}\right)=P\left(G|g_l\cap\overline{B}\right)\cdot P\left(g_l|\overline{B}\right).
	\label{P_Cnd_G_l_nb}
\end{equation}
Since the randomness of choosing the nodes to be attacked causes an independency of the events of choosing a node of some generation and of choosing a node that is not black, then $P\left(g_l|\overline{B}\right)=P\left(g_l\right)$, that is the probability of choosing a node of the $l$'th generation. This probability is the ratio between the number of nodes in the $l$'th generation, which for a CT of $Z$ neighbours is 
$Z\cdot\left(Z-1\right)^{l-1}$, and the total number of nodes in the tree which for CT of $Z$ neighbours is 
$\frac{1}{Z-2}\left[Z\left(Z-1\right)^L-2\right]$. Assuming a very large $L$, this ratio is approximately 
$\left(Z-2\right)\left(Z-1\right)^{-\left(L-l+1\right)}$. Thus we get
\begin{equation}
	P\left(G\cap g_l|\overline{B}\right)=\left[1-\left(1-p_b\right)^l\right]\cdot \left(Z-2\right)\left(Z-1\right)^{-\left(L-l+1\right)}.
\end{equation}  

Following this result, we get that the probability of choosing a grey node given it is not black with no consideration of which generation the node belongs to, is
\begin{align}
	P\left(G|\overline{B}\right)=&\sum_{l=1}^{L}\left[1-\left(1-p_b\right)^l\right]\cdot \left(Z-2\right)\left(Z-1\right)^{-\left(L-l+1\right)}\nonumber\\
	=&\left(Z-2\right)\left(Z-1\right)^{-\left(L+1\right)}\cdot\nonumber\\
	&\qquad\left[\sum_{l=1}^{L}\left(Z-1\right)^l-\sum_{l=1}^{L}\left[\left(Z-1\right)\left(1-p_b\right)^l\right]\right].
\end{align}
We calculate the series and neglect some terms due to the assumption of very large $L$, and get
\begin{equation}
	P\left(G|\overline{B}\right)=1-\frac{\left(Z-2\right)\left(1-p_b\right)^{L+1}}{\left(Z-2\right)-\left(Z-1\right)p_b}.
\end{equation}
Since between the stages $k$ and $k+1$ of an attack, $p_b=\frac{k}{N}$, then between these stages we get
\begin{equation}
	P\left(G|\overline{B}\right)=1-\frac{\left(Z-2\right)\left(1-\frac{k}{N}\right)^{L+1}}{\left(Z-2\right)-\left(Z-1\right)\frac{k}{N}}.
	\label{P_G_nB_gen}
\end{equation}\newline

\textbf{II. Probability of null avalanche during an attack.} \ Equation $\left(\ref{P_G_nB_gen}\right)$ can be interpreted as the probability that on stage $k+1$ of a random attack on the tree, a grey node is chosen to be attacked, i.e. the probability of the event of null avalanche on stage $k+1$. We define an indicator random variable $X_k$ for the event of null avalanche in stage $k+1$, and a random variable $X$ that is the number of null avalanches during an entire attack until the network collapses. Recall that for CT with a very large $L$ and $Z$ neighbours, $p_c=\frac{1}{Z-1}$ (Eq. (\ref{p_c_approx}) above), i.e. an attack on a fraction of $\frac{Z-2}{Z-1}N$ of the network's nodes is required in order to dismantle the network, therefore, $X=\sum_{k=0}^{\frac{Z-2}{Z-1}N-1}X_k$. Accordingly, the expected number of null avalanches during an attack until the network collapses is
\begin{align}
	E\left[X\right]&=\sum_{k=0}^{\frac{Z-2}{Z-1}N-1}E\left[X_k\right]\nonumber\\
	&=\sum_{k=0}^{\frac{Z-2}{Z-1}N-1}\left(1-\frac{\left(Z-2\right)\left(1-\frac{k}{N}\right)^{L+1}}{\left(Z-2\right)-\left(Z-1\right)\frac{k}{N}}\right)\nonumber\\
	&=\frac{Z-2}{Z-1}N-\sum_{k=0}^{\frac{Z-2}{Z-1}N-1}\frac{\left(Z-2\right)\left(1-\frac{k}{N}\right)^{L+1}}{\left(Z-2\right)-\left(Z-1\right)\frac{k}{N}}.
	\label{Ex_gen}
\end{align}
We approximate the last expression in the right-hand-side of Eq. $\left(\ref{Ex_gen}\right)$, by the following integral
\begin{equation}
	\int_{0}^{\frac{Z-2}{Z-1}N-1}\frac{\left(Z-2\right)\left(1-\frac{k}{N}\right)^{L+1}}{\left(Z-2\right)-\left(Z-1\right)\frac{k}{N}}dk.
	\label{integral_approximation_gen}
\end{equation}
We calculate now the integral in Eq. $\left(\ref{integral_approximation_gen}\right)$. For this we use some tools from combinatorics and probability theories as follows, until we get the last expression presented below in Eq. $\left(\ref{Eq_binomial_sum_final_gen}\right)$. We present here the stages of the calculations in general outlines only. For detailed calculations and explanations see the SI. 

By substituting $u=\left(Z-2\right)-\left(Z-1\right)\frac{k}{N}$ and performing some algebraic operations, the integral in Eq. $\left(\ref{integral_approximation_gen}\right)$ is transformed to the following integral
\begin{equation}
	-\frac{\left(Z-2\right)N}{\left(Z-1\right)^{L+2}}\int_{Z-2}^{\frac{Z-1}{N}}\frac{\left(1+u\right)^{L+1}}{u}du.
\end{equation}
Performing the binomial formula of $(1+u)^{L+1}=\sum_{m=0}^{L+1}{L+1 \choose m}u^m$, yields
\begin{equation}
	-\frac{\left(Z-2\right)N}{\left(Z-1\right)^{L+2}}\sum_{m=0}^{L+1}{L+1 \choose m}\int_{Z-2}^{\frac{Z-1}{N}}u^{m-1}du.
	\label{integral_bin_formula_gen}
\end{equation}
Each value of $m$ in the summation in Eq. $\left(\ref{integral_bin_formula_gen}\right)$, yields a definite integral in Eq. $\left(\ref{integral_bin_formula_gen}\right)$ in which the antiderivative of the integrand is a polynomial in $u$, except $m=0$ that yields a definite integral of the integrand $\frac{1}{u}$ that its antiderivative is the logarithm of $u$. Therefore, the expression of $m=0$ is negligible relative to the other expressions of $m\neq0$, so we neglect it and calculate the summation from $m=1$ only. Accordingly, by solving the integral in Eq. $\left(\ref{integral_bin_formula_gen}\right)$ we get
\begin{equation}
	-\frac{\left(Z-2\right)N}{\left(Z-1\right)^{L+2}}\sum_{m=1}^{L+1}{L+1 \choose m}\frac{1}{m}\left[\left(\frac{Z-1}{N}\right)^m-\left(Z-2\right)^m\right],\nonumber
\end{equation}
Since the assumption of a very large $N$, We neglect the term of $\left(\frac{Z-1}{N}\right)^m$ which is much smaller than the term $\left(Z-2\right)^m$, and get
\begin{equation}
	\frac{\left(Z-2\right)N}{\left(Z-1\right)^{L+2}}\sum_{m=1}^{L+1}\left(Z-2\right)^m{L+1 \choose m}\frac{1}{m}.
	\label{integral_largeN_gen}
\end{equation}
Substituting in Eq. $\left(\ref{integral_largeN_gen}\right)$ the approximated value of $N$ for CT with a very large number of layers $L$ that is $\frac{Z}{Z-2}\left(Z-1\right)^L$, gives
\begin{equation}
	\frac{Z}{\left(Z-1\right)^2}\sum_{m=1}^{L+1}\left(Z-2\right)^m{L+1 \choose m}\frac{1}{m}.
	\label{Eq_binomial_sum_gen}
\end{equation}
By applying the binomial coefficients identities $\sum_{j=1}^{L+1}\frac{1}{j}{j \choose m}={L+1 \choose m}\frac{1}{m}$, Eq. $\left(\ref{Eq_binomial_sum_gen}\right)$ gets the formula 
\begin{equation}
	\frac{Z}{\left(Z-1\right)^2}\sum_{j=1}^{L+1}\frac{1}{j}\sum_{m=1}^{L+1}{j \choose m}\left(Z-2\right)^m.
	\label{Eq_binomial_sum_2_gen}
\end{equation}
Since ${j \choose m}=0$ when $m>j$, then on the second summation in Eq. $\left(\ref{Eq_binomial_sum_2_gen}\right)$ we determine the upper limit to be $j$ only. Also $\left(Z-1\right)^j=\sum_{m=0}^{j}{j \choose m}\left(Z-2\right)^m=1+\sum_{m=1}^{j}{j \choose m}\left(Z-2\right)^m$. Substituting it in Eq. $\left(\ref{Eq_binomial_sum_2_gen}\right)$ gives the following formula
\begin{equation}
	\frac{Z}{\left(Z-1\right)^2}\sum_{j=1}^{L+1}\frac{\left(Z-1\right)^j-1}{j}.
	\label{Eq_binomial_sum_2_gen2}
\end{equation}
Changing the variable $j$ to $L-j+1$, applying the identity $\frac{1}{L-j+1}=
\frac{1}{L+1}\sum_{m=0}^{\infty}\frac{j^m}{\left(L+1\right)^m}$ and ignoring terms that are relatively very small, approximates 
Eq. $\left(\ref{Eq_binomial_sum_2_gen2}\right)$ to
\begin{equation}
	Z\left(Z-1\right)^{L-1}\sum_{m=0}^{\infty}\frac{1}{\left(L+1\right)^{m+1}}\sum_{j=0}^{\infty}\frac{j^m}{\left(Z-1\right)^j}.
	\label{Eq_binomial_sum_3_gen}	
\end{equation}
Considering again the size $N$ of CT with $Z$ neighbours and a very large $L$ that is $\frac{Z}{Z-2}\left(Z-1\right)^L$ , gives that $Z\left(Z-1\right)^{L-1}=\frac{Z-2}{Z-1}N$. Substituting it in Eq. $\left(\ref{Eq_binomial_sum_3_gen}\right)$ gives
\begin{equation}
	\frac{Z-2}{Z-1}N\sum_{m=0}^{\infty}\frac{1}{\left(L+1\right)^{m+1}}\sum_{j=0}^{\infty}\frac{j^m}{\left(Z-1\right)^j}.
	\label{Eq_binomial_sum_4_gen}	
\end{equation}
Considering the series $\sum_{j=0}^{\infty}\frac{j^0}{\left(Z-1\right)^j}=\frac{Z-1}{Z-2}$, $\sum_{j=0}^{\infty}\frac{j^1}{\left(Z-1\right)^j}=\frac{Z-1}{\left(Z-2\right)^2}$ and $\sum_{j=0}^{\infty}\frac{j^2}{\left(Z-1\right)^j}=\frac{Z\left(Z-1\right)}{\left(Z-2\right)^3}$, we get the following first terms of the asymptotic expansion of Eq. $\left(\ref{Eq_binomial_sum_4_gen}\right)$
\begin{align}
	&\frac{Z-2}{Z-1}N\cdot\nonumber\\
	&\left(\frac{\frac{Z-1}{Z-2}}{L+1}+\frac{\frac{Z-1}{\left(Z-2\right)^2}}{\left(L+1\right)^2}+\frac{\frac{Z\left(Z-1\right)}{\left(Z-2\right)^3}}{\left(L+1\right)^3}+\mathcal{O}\left(\frac{1}{\left(L+1\right)^4}\right)\right),\nonumber
\end{align} 
that is
 \begin{align}
 	N\left(\frac{1}{L+1}+\frac{\frac{1}{Z-2}}{\left(L+1\right)^2}+\frac{\frac{Z}{\left(Z-2\right)^2}}{\left(L+1\right)^3}+\mathcal{O}\left(\frac{1}{\left(L+1\right)^4}\right)\right)
 \end{align}
Changing the base of the expansion to $L$, yields
\begin{equation}
	N\left(\frac{1}{L}+\frac{\frac{3-Z}{Z-2}}{L^2}+\frac{\frac{Z^2-5Z+8}{2\left(Z-2\right)^2}}{L^3}+\mathcal{O}\left(\frac{1}{L^4}\right)\right).
	\label{Eq_binomial_sum_5_gen}
\end{equation}
Since the assumption of a very large $L$, we neglect the terms of $\mathcal{O}\left(\frac{1}{L^2}\right)$ in Eq. $\left(\ref{Eq_binomial_sum_5_gen}\right)$, and consider only the first term of $\frac{N}{L}$. Therefore, we get the following final approximated expression for the integral in Eq. $\left(\ref{integral_approximation_gen}\right)$ as follows
\begin{equation}
	\int_{0}^{\frac{Z-2}{Z-1}N-1}\frac{\left(Z-2\right)\left(1-\frac{k}{N}\right)^{L+1}}{\left(Z-2\right)-\left(Z-1\right)\frac{k}{N}}dk
	\approx\frac{N}{L}.
	\label{Eq_binomial_sum_final_gen}
\end{equation}
Substituting Eq. $\left(\ref{Eq_binomial_sum_final_gen}\right)$ into Eq. $\left(\ref{Ex_gen}\right)$ gives for the expected number of null avalanches during an entire attack as follows
\begin{equation}
	E\left[X\right]=\frac{Z-2}{Z-1}N-\frac{N}{L}.
	\label{expectation_X_gen}
\end{equation}
Assume a random choice of one attack stage from the set of $\frac{Z-2}{Z-1}N$ attack stages. We denote by $S_0$ the event of null avalanche that occurs due to the node removal in the chosen stage. Accordingly, the probability of $S_0$ is the ratio of the expectation of the number of null avalanches during an entire attack $E\left[X\right]$, calculated above in Eq. $\left(\ref{expectation_X_gen}\right)$, and the number of the attack stages $\frac{Z-2}{Z-1}N$. Therefore we get
\begin{equation}
	P\left(S_0\right)=1-\frac{\frac{Z-1}{Z-2}}{L}.
	\label{probability_nullaval}
\end{equation}

\subsubsection{Probability of complete avalanche}
\textbf{I. Definitions.} \ We state the following definitions: for this part, the event of choosing randomly a node of wight mode and of the $l$'th generation is denoted $W_l$. By $W^1_l$ we note the same event as $W_l$, plus that the chosen node is related as son no. 1 of its father, and so is the event $W^2_l$ for a node related as son no. 2 of its father,  $W^3_l$ for a node related as son no. 3 of its father and so on ... until  $W^{Z-1}_l$ for a node related as son no. $Z-1$ of its father

A `complete cluster' is a set of nodes consists of a father and all its descendants, where all of them are in wight mode. The father is named the `head' of the complete cluster. If a father of a complete cluster is attacked, than a complete avalanche occurs. The event of choosing randomly a node that is a head of a complete cluster is denoted $C$. By $C^1$ we note the same event as $C$ plus that the chosen node is related as son no. 1 of its father, and so is the event $C^2$ for a node related as son no. 2 of its father, the event $C^3$ for a node related as son no. 3 of its father, and so on..., until the event $C^{Z-1}$ for a node related as son no. $Z-1$ of its father.\newline

\textbf{II. Conditional probability of complete avalanche of size $\frac{\left(Z-1\right)^m-1}{Z-2} \ (m=1,2,3,\dots,L)$ in the $k+1$ stage of an attack.} \ We define $q_l$ as the probability of choosing randomly a node that is a head of complete cluster, given that this node is of wight mode and of the $l$'th generation, i.e. $q_l=P(C|W_l)$.

For simplicity, from now on we develop the following principles for CT with $Z=3$ neighbours, and later we would generalize the results to a CT with any number of neighbours $Z$. For a CT with $Z=3$, the probability $q_l$ can also be related to the events related to the two sons of the chosen node, as follows
\begin{equation}
	q_l=P\left[\left(C^1\cap W^1_{l+1}\right)\cap \left(C^2\cap W^2_{l+1}\right)|W_l\right].
	\label{q_l_firstterm}
\end{equation}
Since the independency of the sons, the previous expression can be written as follows
\begin{equation}
	q_l=P\left[\left(C^1\cap W^1_{l+1}\right)|W_l\right]\cdot P\left[\left(C^2\cap W^2_{l+1}\right)|W_l\right].
	\label{q_l_independency}
\end{equation}
We develop now the term $P\left[\left(C^1\cap W^1_{l+1}\right)|W_l\right]$. According to the conditional probability formula $P\left(A\cap B|C\right)=P\left(A|B\cap C\right)\cdot P\left(B|C\right)$, we get for this term
\begin{equation}
	P\left[\left(C^1\cap W^1_{l+1}\right)|W_l\right]=P\left(C^1|W^1_{l+1}\cap W_l\right)\cdot P\left(W^1_{l+1}|W_l\right).
\end{equation}
Since $W^1_{l+1}\subset W_l$, this can be written as follows
\begin{align}
	&P\left(C^1|W^1_{l+1}\right)\cdot P\left(W^1_{l+1}|W_l\right)\nonumber\\
	=&q_{l+1}(1-p_b).
	\label{q_l_partial_recursion_relation}
\end{align}
Since $P\left[\left(C^1\cap W^1_l\right)|W_l\right]$=$P\left[\left(C^2\cap W^2_l\right)|W_l\right]$, $q_l$ in Eq. $\left(\ref{q_l_independency}\right)$ could be written as the following recursion relation
\begin{equation}
	q_l=\left(1-p_b\right)^2 q_{l+1}^2,
	\label{q_l_recursion_relation}
\end{equation}
where $l=1,2,3,\ \dots\ ,L-1,L$. 
Referring to a complete cluster of size $1$ which is a wight node from the surface of the tree i.e. the last tree layer $L$, the event of choosing a head of complete cluster given that this node is of the $L$'th generation and of wight mode, is a sure event with probability $1$. So, combined with the recursion relation of Eq. $\left(\ref{q_l_recursion_relation}\right)$, we also have the boundary condition $q_L=1$. 

The result of Eq. $\left(\ref{q_l_recursion_relation}\right)$ can be now generalized to a CT with any number $Z$ of neighbours, by applying the same rationales of Eqs. $\left(\ref{q_l_firstterm}\right)$ -- $\left(\ref{q_l_partial_recursion_relation}\right)$ were presented for a CT with $Z=3$. We accordingly get a recursion relation with a boundary condition for the general case of CT with $Z$ neighbours as follows
\begin{align}
 	&q_l=\left(1-p_b\right)^{Z-1} q_{l+1}^{Z-1}.\nonumber\\
 	&q_L=1.
 	\label{q_l_recursion_relation_gen}
 \end{align}
Using this we get the following formulas
\begin{align}
	&q_{L-1}=\left(1-p_b\right)^{Z-1}q_L^{Z-1}=\left(1-p_b\right)^{Z-1}\nonumber\\
	&q_{L-2}=\left(1-p_b\right)^{Z-1}q_{L-1}^{Z-1}=\left(1-p_b\right)^{\left(Z-1\right)^2+\left(Z-1\right)},\nonumber
\end{align}
and in general
\begin{align}
	q_{L-m+1}&=\left(1-p_b\right)^{\sum_{j=1}^{m-1}\left(Z-1\right)^j}\nonumber\\
	&=\left(1-p_b\right)^{\frac{1}{Z-2}\left(\left(Z-1\right)^m-Z+1\right)}.
\end{align}
Taking in consideration that between the stages $k$ and $k+1$ of an attack, $p_b=\frac{k}{N}$, we can state a general term for the probability of
attacking randomly in the $k+1$ stage of an attack a node that is a head of complete cluster, given that this node is of wight mode and of the $L-m+1$ generation, that is the probability of attacking in the $k+1$ stage of an attack a node that causes a complete avalanche of size $\frac{\left(Z-1\right)^m-1}{Z-2}$ given that this node is of wight mode, as follows
\begin{equation}
	q_{L-m+1}=\left(1-\frac{k}{N}\right)^{\frac{1}{Z-2}\left(\left(Z-1\right)^m-Z+1\right)},
	\label{q_L-m_final_gen}
\end{equation}
where $m=1,2,3\ \dots \ ,L$.\\

\textbf{III. Probability of complete avalanche of size $\frac{\left(Z-1\right)^m-1}{Z-2} \ (m=1,2,3,\dots,L)$ in the $k+1$ stage of an attack.} \ 
We define $Q_{L-m+1}$ as the probability of choosing randomly a node that is a head of complete cluster and this node is of wight mode and of the generation $L-m+1$, given that node is not black, i.e. $Q_{L-m+1}=P\left(C\cap W\cap g_{L-m+1}|\overline{B}\right)$. Applying the conditional probability formula, we get 
\begin{align}
	&Q_{L-m+1}\nonumber\\
	=&P\left(C|W\cap g_{L-m+1}\cap\overline{B}\right)\cdot P\left(W|g_{L-m+1}\cap\overline{B}\right)\cdot\nonumber\\
	 &\ \cdot P\left(g_{L-m+1}|\overline{B}\right).
	 \label{Q_L-m_initial}
\end{align}
Since $W\subset \overline{B}$, then $W\cap g_{L-m+1}\cap\overline{B}=W\cap g_{L-m+1}$. Therefore, $P\left(C|W\cap g_{L-m+1}\cap\overline{B}\right)=P\left(C|W\cap g_{L-m+1}\right)=q_{L-m+1}$. According to Eq. $\left(\ref{P_W_l_nB}\right)$, $P\left(W|g_{L-m+1}\cap\overline{B}\right)=\left(1-p_b\right)^{L-m+1}$. Since the randomness of the attack, the non-black network's nodes are distributed between the network's layers with the same proportions as between the layers' sizes, causes that $P\left(g_{L-m+1}|\overline{B}\right)=P\left(g_{L-m+1}\right)$. The probability $P\left(g_{L-m+1}\right)$ is the ratio between the size of the layer $L-m+1$ which is $Z\left(Z-1\right)^{L-m}$ and the network size which is approximately $\frac{Z}{Z-2}\left(Z-1\right)^L$. Substitute all of these considerations into Eq. $\left(\ref{Q_L-m_initial}\right)$, and taking again in consideration that between the stages $k$ and $k+1$ of an attack, $p_b=\frac{k}{N}$, gives
\begin{align}
 	Q_{L-m+1}&=\left(1-\frac{k}{N}\right)^{\frac{1}{Z-2}\left(\left(Z-1\right)^m-Z+1\right)+L-m+1}\cdot\nonumber\\
	&\left(Z-2\right)\left(\frac{1}{Z-1}\right)^m.	
 \end{align}
where $m=1,2,\ \dots \ ,L$.\\

By definition, $Q_{L-m+1}$ is the probability of attacking randomly in the $k+1$ stage of an attack a node that causes a complete avalanche of size $\frac{\left(Z-1\right)^m-1}{Z-2}$.\\

\textbf{IV. Probability of complete avalanche of size $\frac{\left(Z-1\right)^m-1}{Z-2}$ during an entire attack.} \ We define again an indicator random variable $X_k$ for the event of a complete avalanche of size $\frac{\left(Z-1\right)^m-1}{Z-2}$ in stage $k+1$ of an attack, and a random variable $X$ that is the number of complete avalanches of size $\frac{\left(Z-1\right)^m-1}{Z-2}$ during an entire attack until the network collapses, i.e. $X=\sum_{k=0}^{\frac{Z-2}{Z-1}N-1}X_k$. Therefore, the expected number of complete avalanches of size $\frac{\left(Z-1\right)^m-1}{Z-2}$ during an attack until the network collapses is
\begin{align}
	E\left[X\right]=\sum_{k=0}^{\frac{Z-2}{Z-1}N-1}&E\left[X_k\right]=\nonumber\\
	\sum_{k=0}^{\frac{Z-2}{Z-1}N-1}\left(1-\frac{k}{N}\right)&^{\frac{1}{Z-2}\left(\left(Z-1\right)^m-Z+1\right)+L-m+1}\cdot\nonumber\\
	&\left(Z-2\right)\left(\frac{1}{Z-1}\right)^m.
\end{align}
To calculate this term, we approximate it by the following integral
\begin{align}
	E\left[X\right]&=\left(Z-2\right)\left(\frac{1}{Z-1}\right)^m\nonumber\\
	&\int_0^{\frac{Z-2}{Z-1}N-1}\left(1-\frac{k}{N}\right)^{\frac{1}{Z-2}\left(\left(Z-1\right)^m-Z+1\right)+L-m+1}dk.
\end{align}
By changing variables with the formula $u=1-\frac{k}{N}$ and performing some algebraic operations, we get
\begin{align}
	E\left[X\right]&=\left(Z-2\right)\left(\frac{1}{Z-1}\right)^m\nonumber\\
	&\int_1^{\frac{1}{Z-1}+\frac{1}{N}}u^{\frac{1}{Z-2}\left(\left(Z-1\right)^m-Z+1\right)+L-m+1}\cdot\left(-N\right)du.
\end{align}
 Solving the integral and neglecting terms which are relatively very small, yields
 \begin{align}
 	E[X]&=\left(Z-2\right)\left(\frac{1}{Z-1}\right)^mN\cdot\nonumber\\
 	&\frac{1}{\frac{1}{Z-2}\left(\left(Z-1\right)^m-Z+1\right)+L+2-m}.
 	\label{Expected_specific_avalanche_gen}
 \end{align} \\
  We rewrite Eq. $\left(\ref{Expected_specific_avalanche_gen}\right)$ using an exponential term, and get
 \begin{align}
	E\left[X\right]=N&\left(Z-2\right)\left(\frac{1}{Z-1}\right)^m\frac{1}{L}\cdot\nonumber\\
	&e^{-ln\left(1+\frac{\frac{\left(Z-1\right)^m}{Z-2}-\frac{Z-1}{Z-2}+2-m}{L}\right)}.
	\label{Expected_specific_avalanche_gen_rearr}
\end{align}
Assume again a random choice of one attack stage from the set of $\frac{Z-2}{Z-1}N$ attack stages. We denote by $S_m$ the event of a complete avalanche of size $\frac{\left(Z-1\right)^m-1}{Z-2}$ that occurs due to the node removal in the chosen stage. Accordingly, the probability of $S_m$ is the ratio of the expectation of the number of complete avalanches of size $\frac{\left(Z-1\right)^m-1}{Z-2}$ during an entire attack $E\left[X\right]$, calculated above in Eq. $\left(\ref{Expected_specific_avalanche_gen_rearr}\right)$, and the number of the attack stages $\frac{Z-2}{Z-1}N$. Therefore we get
\begin{align}
	P\left(S_m\right)=&\left(\frac{1}{Z-1}\right)^{m-1}\frac{1}{L}\cdot\nonumber\\
	&e^{-ln\left(1+\frac{\frac{\left(Z-1\right)^m}{Z-2}-\frac{Z-1}{Z-2}+2-m}{L}\right)}.
	\label{probability_Sm_gen}
\end{align}
Due to the dominance of the term $\left(Z-1\right)^m$ than the other terms in the exponent of Eq. $\left(\ref{probability_Sm_gen}\right)$, the term $e^{-ln\left(1+\frac{\frac{\left(Z-1\right)^m}{Z-2}-\frac{Z-1}{Z-2}+2-m}{L}\right)}$ approaches $1$ when $m$ is small relative to $\log_{Z-1}L$ and approaches $0$ when $m$ is great relative to $\log_{Z-1}L$, with a very narrow region around $\log_{Z-1}L$ where the transition between $1$ and $0$ occurs. Therefore, we approximate this term to be $1$ for values of $m$ that are less or equal to $\log_{Z-1}L$, and to be $0$ for values of $m$ that are greater than $\log_{Z-1}L$. Applying this approximation in Eq. $\left(\ref{probability_Sm_gen}\right)$, gives the following distribution of a complete avalanche of size $\frac{\left(Z-1\right)^m-1}{Z-2}$ $\left(m=1,2,3,\ldots,L\right)$ as follows
\begin{equation}
	P\left(S_m\right)=
	\begin{cases}
		\frac{1}{L}\cdot\frac{1}{\left(Z-1\right)^{m-1}} & m\leq \log_{Z-1}L\\
		0 & m>\log_{Z-1}L\\
	\end{cases}  
	\label{comp_aval_distrib_gen}     
\end{equation}

\textbf{V. Probability of complete avalanche of any size during an entire attack.} \ Assume a random choice of one attack stage from the set of $\frac{Z-2}{Z-1}N$ attack stages. We denote by $S_C$ the event of complete avalanche that occurs due to the node removal in the chosen stage. The probability of $S_C$ is the summation of the probabilities of the events $S_m$. Using Eq.$\left(\ref{comp_aval_distrib_gen}\right)$ accordingly gives 
\begin{equation}
	P\left(S_C\right)=\sum_{m=1}^{\log_{Z-1}L}\frac{1}{L}\left(\frac{1}{Z-1}\right)^{m-1}.\nonumber
\end{equation}
Calculation of the previous series gives the following
\begin{equation}
	P\left(S_C\right)=\frac{\frac{Z-1}{Z-2}}{L}\left(1-\frac{1}{L}\right).
\end{equation}
Under the assumption of a very large $L$, we neglect the term of $\frac{1}{L}$ in the brackets, and get a final expression as follows
\begin{equation}
	P\left(S_C\right)=\frac{\frac{Z-1}{Z-2}}{L}.
	\label{probability_compaval}
\end{equation}

\subsubsection{Summary of theory}
The conclusions that are obtained from the considerations and calculations detailed above, of the distribution of the avalanche's sizes in a random attack on CT with $Z$ neighbours, are as follows: Under the assumption of a very large $L$, the probability of the event of null avalanche in a randomly chosen stage of an attack is (Eq. $\left(\ref{probability_nullaval}\right)$)
\begin{equation}
	P\left(S_0\right)=1-\frac{\frac{Z-1}{Z-2}}{L}.
	\label{Eq_ps0_final}
\end{equation}
The probability of the event of a complete avalanche of size $\frac{\left(Z-1\right)^m-1}{Z-2}$ $\left(m=1,2,3,\dots,L\right)$ in a randomly chosen stage of an attack is (Eq. $\left(\ref{comp_aval_distrib_gen}\right)$)
\begin{equation}
	P\left(S_m\right)=
	\begin{cases}
		\frac{1}{L}\cdot\frac{1}{\left(Z-1\right)^{m-1}} & m\leq \log_{Z-1}L\\
		0 & m>\log_{Z-1}L\\
	\end{cases}  
\end{equation}
Under the assumption of a very large $L$, the probability of the event of a complete avalanche of any size in a randomly chosen stage of an attack is (Eq. $\left(\ref{probability_compaval}\right)$)
\begin{equation}
	P\left(S_C\right)=\frac{\frac{Z-1}{Z-2}}{L}.
	\label{Eq_psc_final}
\end{equation}
%%%%%%%%%%% end generalization %%%%%%%%%%

\begin{figure*}[htbp]
	\begin{center}
		\begin{tabular}{cc}
			%\hspace{-6.1em} \includegraphics[scale=0.3]{fig2a.eps}\hspace{-2.4em} &
			%\includegraphics[scale=0.3]{fig2b.eps} \\
			\includegraphics[scale=0.4]{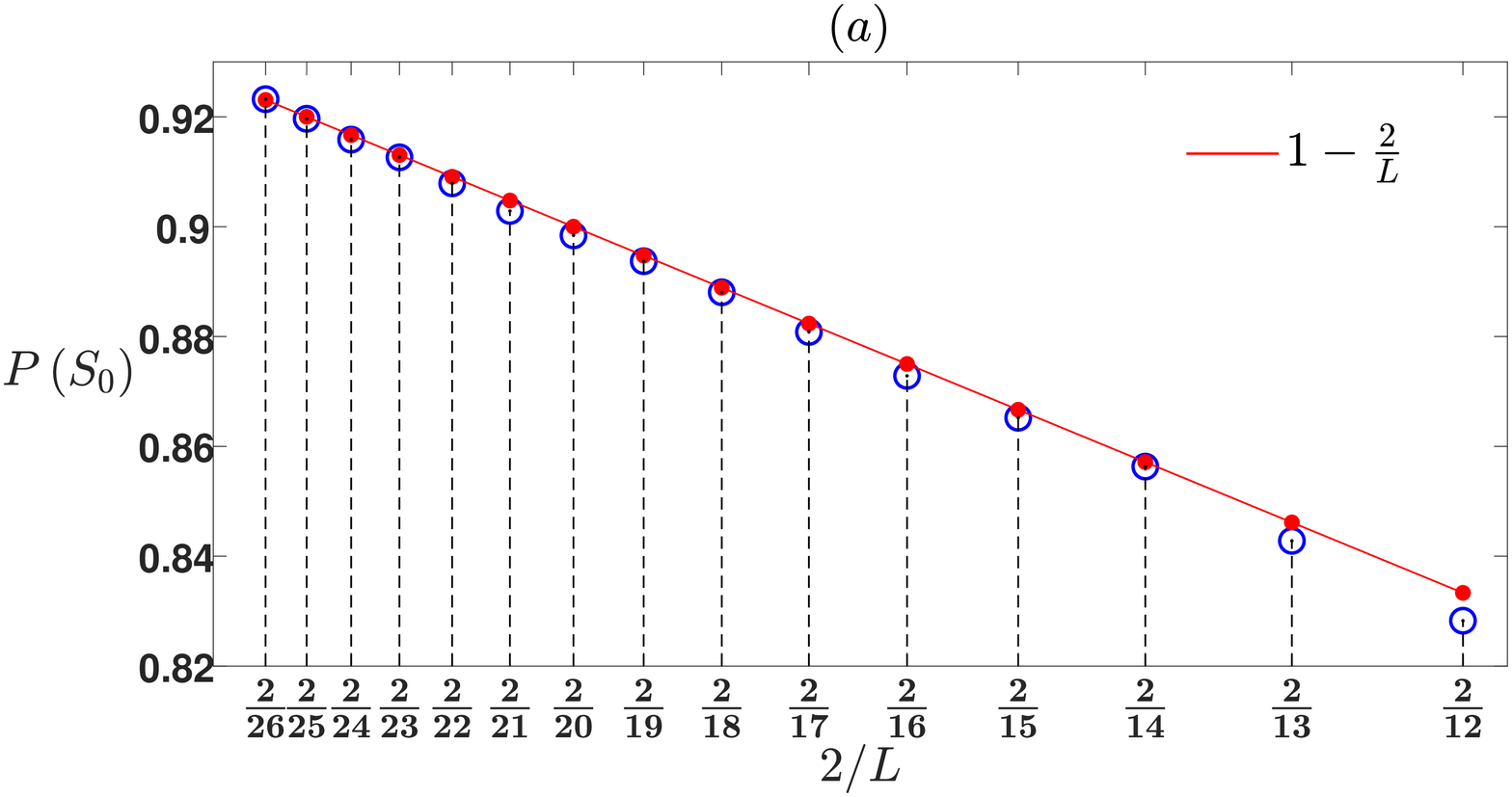}\\
			\includegraphics[scale=0.4]{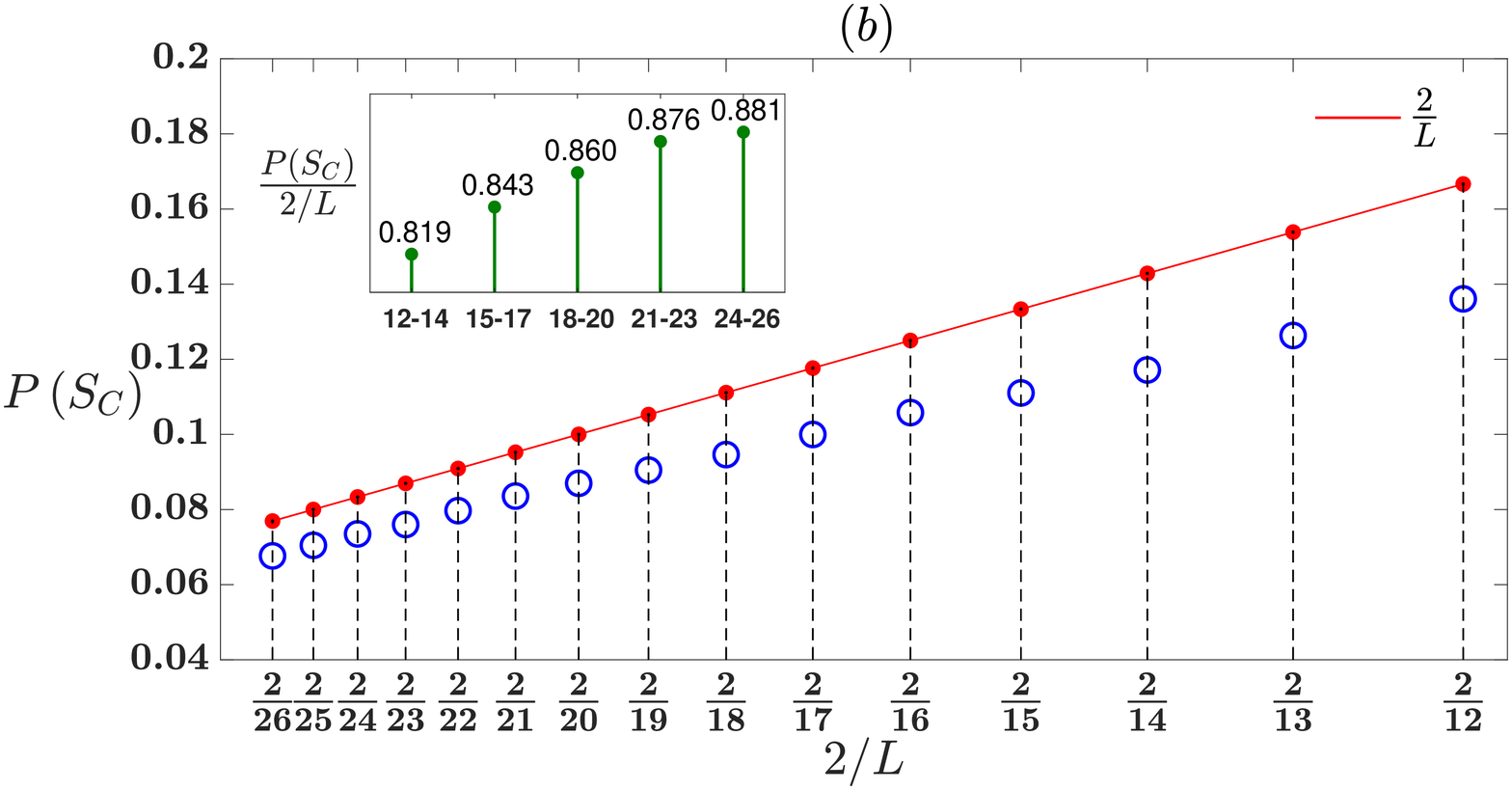}
		\end{tabular}
	\end{center}
	\caption{$\left[\left(a\right),\left(b\right)\right]$ Presentation of avalanche-size distribution for CT with $Z=3$ neighbours. $\left(a\right)$ Presentation of $P\left(S_0\right)$ as predicted by theory and as resulted by simulations, vs $\frac{2}{L}$. Scale marks of x-axis are $\frac{2}{26}, \frac{2}{25}, \frac{2}{24}\ldots, \frac{2}{12}$ refer to CT with $L=26,25,24\ldots, 12$ respectively. At the top of the vertical dashed line belongs to each of scale mark of $\frac{2}{L}$, a red dot represents the value of $1-\frac{2}{L}$ predicted by theory, and a blue circle represents the value of $P\left(S_0\right)$ obtained from simulations. Red line is a graph of the line $1-\frac{2}{L}$. $\left(b\right)$ Presentation of $P\left(S_C\right)$ as predicted by theory and as resulted by simulations, vs $\frac{2}{L}$. At the top of the vertical dashed line belongs to each of scale mark of $\frac{2}{L}$, a red dot represents the value of $\frac{2}{L}$ predicted by theory, and a blue circle represents the value of $P\left(S_C\right)$ obtained from simulations. Red line is a graph of the line $\frac{2}{L}$. Inset in panel $\left(b\right)$: Bar graph of mean value of the ratio of $P\left(S_C\right)$ and $\frac{2}{L}$ for five subsets of $L$ values: $12-14, 15-17, 18-20, 21-23$ and $24-26$. Averages were taken over $1000$ realizations.}
	\label{fig:CT_2}  
\end{figure*}

\subsection{Simulations and results}
Figure \ref{fig:CT_2} illustrates the validity of the theory for CT with $Z=3$. Recall that for this CT, the theory predicts as follows (following Eqs. $\left(\ref{Eq_ps0_final}\right)$ -- $\left(\ref{Eq_psc_final}\right)$): Under the assumption of a very large $L$, the probability of the event of null avalanche in a randomly chosen stage of an attack is
\begin{equation}
	P\left(S_0\right)=1-\frac{2}{L}.\nonumber
\end{equation}
The probability of the event of complete avalanche of size $2^m-1$ $\left(m=1,2,3,\dots,L\right)$ in a randomly chosen stage of an attack is
\begin{equation}
	P\left(S_m\right)=
	\begin{cases}
		\frac{1}{L}\cdot\frac{1}{2^{m-1}} & m\leq \log_2L\nonumber\\
		0 & m>\log_2L\nonumber\\
	\end{cases}  
\end{equation}
Under the assumption of a very large $L$, the probability of the event of complete avalanche in a randomly chosen stage of an attack is
\begin{equation}
	P\left(S_C\right)=\frac{2}{L}.\nonumber
\end{equation}
Also for CT with $Z=3$, the number of stages of an attack until the network is dismantled is $\frac{Z-2}{Z-1}N=\frac{N}{2}$.

The figure presents the results of simulations of CT with $Z=3$ and $L=12, 13, 14,\ldots, 26$. In Fig. \ref{fig:CT_2}$\left(a\right)$ the graph presents the probability of null avalanche $P\left(S_0\right)$, as predicted by the theory and as was obtained by simulations, versus $\frac{2}{L}$. The scale marks of x-axis that are $\frac{2}{26}, \frac{2}{25}, \frac{2}{24},\ldots, \frac{2}{12}$ refer to CT with $L=26,25,24,\ldots, 12$, respectively. For each value of $\frac{2}{L}$, at the top of the vertical dashed line belongs to it a red dot represents the value of $1-\frac{2}{L}$ which is the probability $P\left(S_0\right)$ as predicted by the theory, and a blue circle represents the value of $P\left(S_0\right)$ obtained from simulations and which was calculated to be the average of the number of null avalanches in $1000$ realizations were simulated divided by the number of attack stages $\frac{N}{2}$. It is shown that as $L$ increases from $12$ to $26$, i.e $\frac{2}{L}$ decreases from $\frac{2}{12}$ to $\frac{2}{26}$, the blue circles tend to coincide with the red dots. For $L=25$ and $26$, there is indeed a good match between the blue circle and the red dot. That means that as $L$ increases, $P\left(S_0\right)$ as was calculated from simulations, tends to the value predicted by our theory that is $1-\frac{2}{L}$.
In Fig. \ref{fig:CT_2}$\left(b\right)$ the graph presents the probability of complete avalanche $P\left(S_C\right)$, as predicted by the theory and as was obtained by simulations, versus $\frac{2}{L}$. As in panel $\left(a\right)$, for each value of $\frac{2}{L}$, at the top of the vertical dashed line belongs to it a red dot represents the value of $\frac{2}{L}$ which is the value of $P\left(S_C\right)$ as predicted by the theory, and a blue circle represents the value of $P\left(S_C\right)$ obtained from simulations and which again was calculated to be the average of the number of complete avalanches in $1000$ realizations were simulated divided by the attack stages $\frac{N}{2}$. It is shown that as $L$ increases from $12$ to $26$, the blue circles become close to the red dots. This is illustrated more strictly in the inset of Fig. \ref{fig:CT_2}$\left(b\right)$ as follows -- for each of the fifteen CT's were simulated (CT's with $L=12, 13, 14,\ldots, 26$), the ratio of $P\left(S_C\right)$ was obtained from the simulations and $\frac{2}{L}$ was calculated. Then the set of the $L$ values was divided to the five following subsets -- $12-14, 15-17, 18-20, 21-23$ and $24-26$. For each of the five subsets, the mean of the ratio between $P\left(S_C\right)$ and $\frac{2}{L}$ considering the three members of the subset, was calculated. For example, for the subset of the $L$ values $12, 13$ and $14$, the following mean was calculated: $\frac{1}{3}\cdot\left(\frac{P\left(S_C\right)_{12}}{2/12}+\frac{P\left(S_C\right)_{13}}{2/13}+\frac{P\left(S_C\right)_{14}}{2/14}\right)$, where $P\left(S_C\right)_{12}$ is the probability of complete avalanche was obtained from simulations on CT with $L=12$, and so are $P\left(S_C\right)_{13}$ and $P\left(S_C\right)_{14}$. Then a bar graph, that is the in the inset of Fig. \ref{fig:CT_2}$\left(b\right)$, was sketched, where each of the five subsets is represented by one bar whose height is the mean of the ratio of $P\left(S_C\right)$ and $\frac{2}{L}$ calculated for this subset. The value of this ratio is marked on top of the bar. Recall that according to our theory, as $L$ increases, $P\left(S_C\right)$ becomes greater and tends to $\frac{2}{L}$, that is $\frac{P\left(S_C\right)}{2/L}$ becomes greater and tends to $1$. In the bar graph was sketched, it is firstly shown that as $L$ increases between the subsets, the mean of the ratio $\frac{P\left(S_C\right)}{2/L}$ becomes greater, i.e. from $0.819$ for the the subset $12-14$ to $0.881$ for the subset $24-26$, and approaches towards the value $1$ (note that in order to get values that are close to the limit $1$, a network with an enormous size (mathematically tends to infinity) should be simulated). Another evidence of the validity of the theory is related to the prediction of the probability of complete avalanches of size $2^m-1$, to be $\frac{1}{L}\cdot\frac{1}{2^{m-1}}$ if $m\leq \log_2L$ and $0$ if $m>\log_2L$. We take for example the first subset where $L$ varies between $12$ to $14$, and respectively $\log_2L$ varies between $\log_212=3.58$ to $\log_214=3.81$. Accordingly, for this subset the theory predicts that the probability of complete avalanches of size $2^m-1$, is $\frac{1}{L}\cdot\frac{1}{2^{m-1}}$ if $m\leq 3$ and $0$ if $m>3$. With accordance to it, the probability of complete avalanche of any size $P\left(S_C\right)$ is predicted to be $\frac{1}{L}\left(\frac{1}{2^{1-1}}+\frac{1}{2^{2-1}}+\frac{1}{2^{3-1}}\right)=\frac{1.75}{L}$, and $\frac{P\left(S_C\right)}{2/L}$ is predicted to be $0.875$. In the inset of Fig. \ref{fig:CT_2}$\left(b\right)$ it is indeed shown that the simulation's result of $\frac{P\left(S_C\right)}{2/L}$ for this subset is $0.819$, which is $93.6\%$ of the predicted value $0.875$. 
In the same way we take for example the last subset where $L$ varies between $24$ to $26$ and respectively $\log_2L$ varies between $\log_224=4.58$ to $\log_225=4.64$, so for this subset the theory predicts that the probability of complete avalanches of size $2^m-1$, is $\frac{1}{L}\cdot\frac{1}{2^{m-1}}$ if $m\leq 4$ and $0$ if $m>4$. Therefore, the probability of complete avalanche of any size $P\left(S_C\right)$ is predicted to be $\frac{1}{L}\left(\frac{1}{2^{1-1}}+\frac{1}{2^{2-1}}+\frac{1}{2^{3-1}}+\frac{1}{2^{4-1}}\right)=\frac{1.875}{L}$, and accordingly $\frac{P\left(S_C\right)}{2/L}$ is predicted to be $0.9375$. In the inset of Fig. \ref{fig:CT_2}$\left(b\right)$ it is indeed shown that the simulation's result of $\frac{P\left(S_C\right)}{2/L}$ for this subset is $0.881$, which is $94\%$ of the predicted value $0.9375$. Similarly these are the results of the other subsets too.

\begin{figure*}[htbp]
	\begin{center}
		\begin{tabular}{cc}
			%\hspace{-6.1em} \includegraphics[scale=0.3]{fig2a.eps}\hspace{-2.4em} &
			%\includegraphics[scale=0.3]{fig2b.eps} \\
			\includegraphics[scale=0.4]{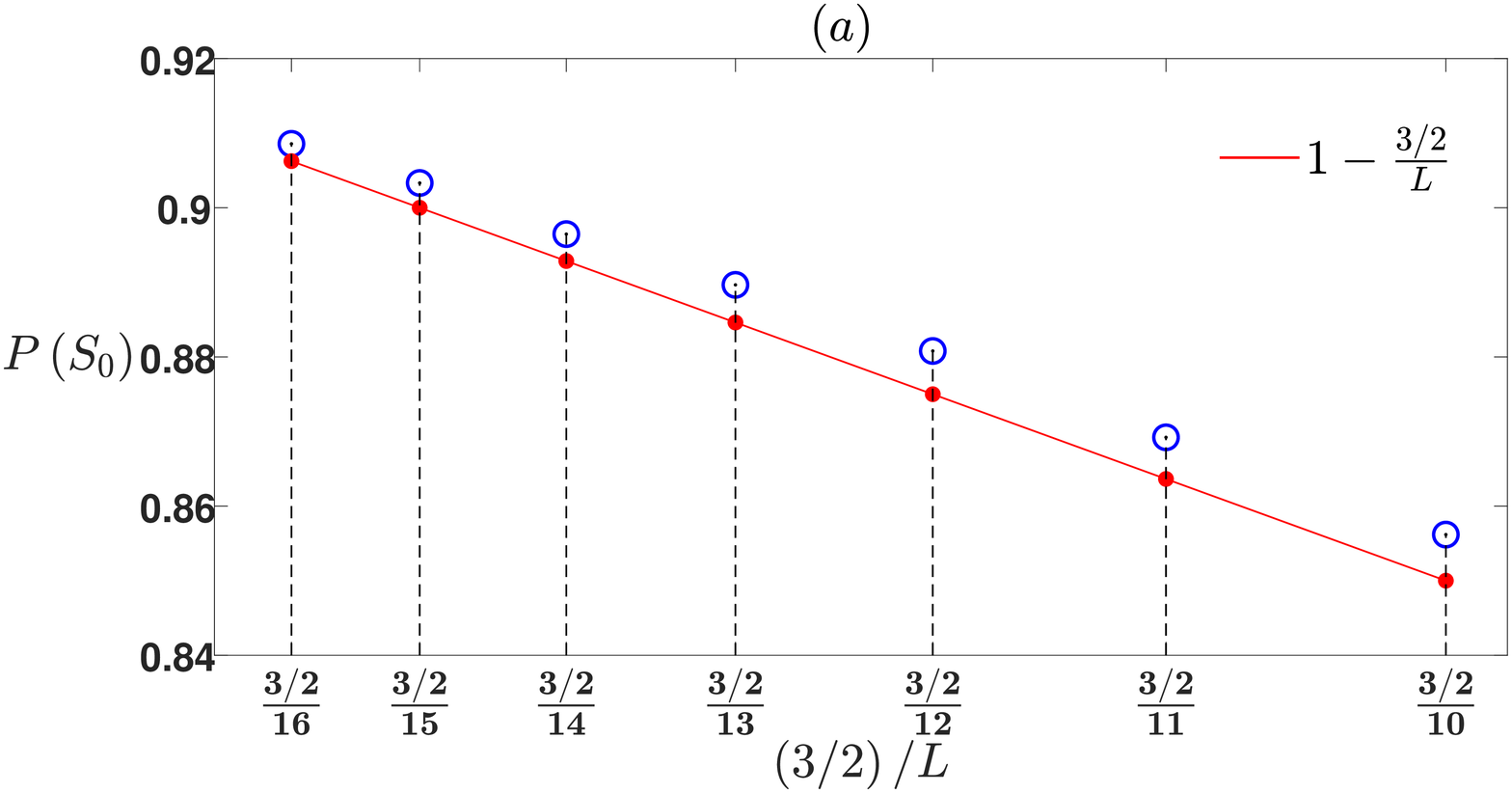}\\
			\includegraphics[scale=0.4]{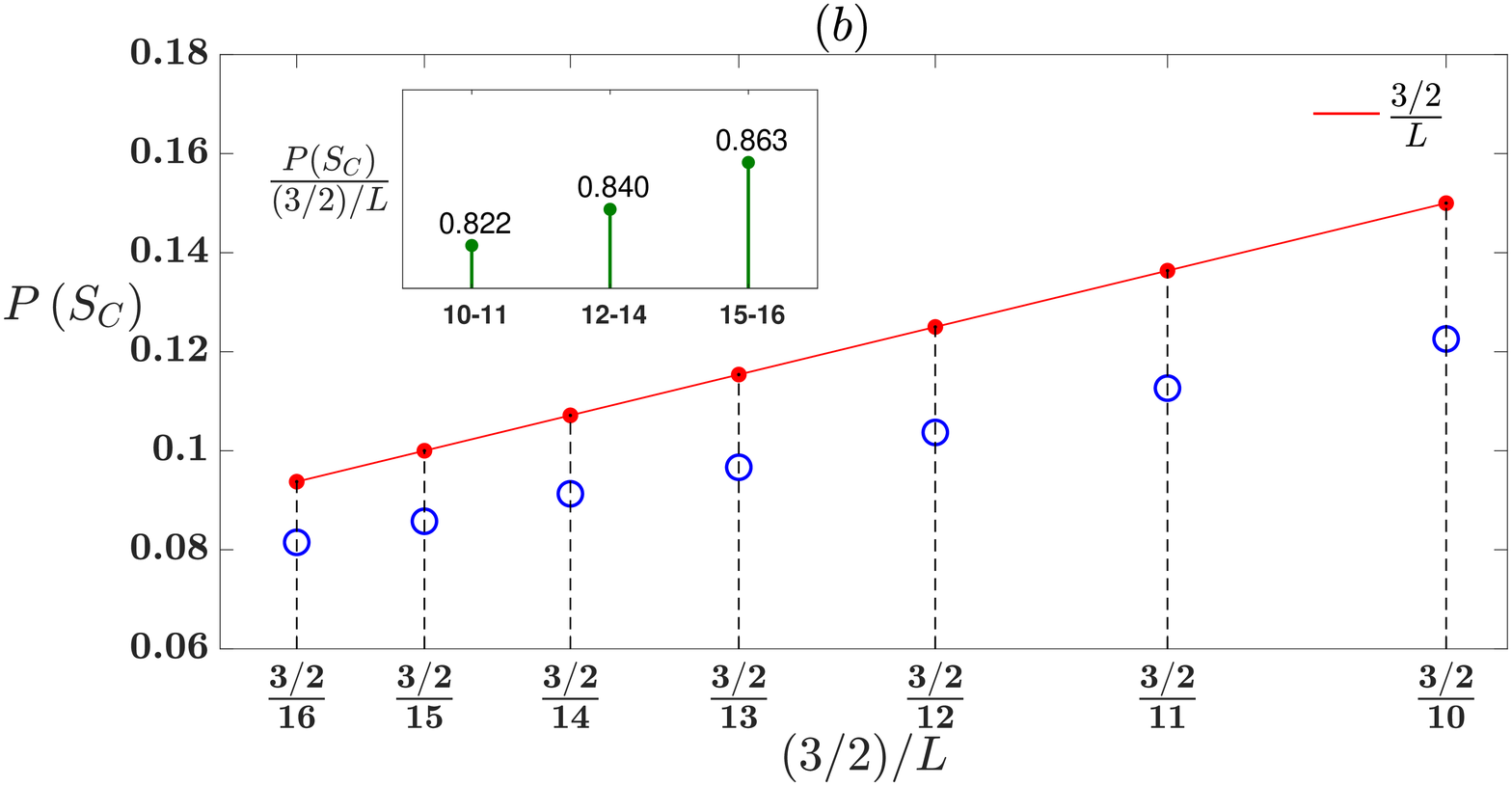}
		\end{tabular}
	\end{center}
	\caption{$\left[\left(a\right),\left(b\right)\right]$ Presentation of avalanche-size distribution for CT with $Z=4$ neighbours. $\left(a\right)$ Presentation of probability of null avalanche $P\left(S_0\right)$ as predicted by theory and as resulted by simulations, vs $\frac{3/2}{L}$. Scale marks of x-axis are $\frac{3/2}{16}, \frac{3/2}{15}, \frac{3/2}{14}\ldots, \frac{3/2}{10}$ refer to CT with $L=16,15,14\ldots, 10$ respectively. At the top of the vertical dashed line belongs to each of scale mark of $\frac{3/2}{L}$, a red dot represents the value of $1-\frac{3/2}{L}$ predicted by theory, and a blue circle represents the value of $P\left(S_0\right)$ obtained from simulations. Red line is a graph of the line $1-\frac{3/2}{L}$. $\left(b\right)$ Presentation of probability of complete avalanche $P\left(S_C\right)$ as predicted by theory and as resulted by simulations, vs $\frac{3/2}{L}$. At the top of the vertical dashed line belongs to each of scale mark of $\frac{3/2}{L}$, a red dot represents the value of $\frac{3/2}{L}$ predicted by theory, and a blue circle represents the value of $P\left(S_C\right)$ obtained from simulations. Red line is a graph of the line $\frac{3/2}{L}$. Inset in panel $\left(b\right)$: Bar graph of mean value of the ratio of $P\left(S_C\right)$ and $\frac{3/2}{L}$ for three subsets of $L$ values: $10-11, 12-14$ and $15-16$. Averages were taken over $1000$ realizations.}
	\label{fig:CT_Z4}  
\end{figure*}

Figure \ref{fig:CT_Z4} is another illustration of the validity of the theory, for CT with $Z=4$. For this CT, the theory again predicts as follows: Under the assumption of a very large $L$, the probability of the event of null avalanche in a randomly chosen stage of an attack is
\begin{equation}
	P\left(S_0\right)=1-\frac{3/2}{L}.\nonumber
\end{equation}
The probability of the event of complete avalanche of size $\frac{3^m-1}{2}$ $\left(m=1,2,3,\dots,L\right)$ in a randomly chosen stage of an attack is
\begin{equation}
	P\left(S_m\right)=
	\begin{cases}
		\frac{1}{L}\cdot\frac{1}{3^{m-1}} & m\leq \log_3L\nonumber\\
		0 & m>\log_3L\nonumber\\
	\end{cases}  
\end{equation}
Under the assumption of a very large $L$, the probability of the event of complete avalanche in a randomly chosen stage of an attack is
\begin{equation}
	P\left(S_C\right)=\frac{3/2}{L}.\nonumber
	\label{Eq_psc-final}
\end{equation}
Also for CT with $Z=4$, the number of stages of an attack until the network is dismantled is $\frac{Z-2}{Z-1}N=\frac{2}{3}N$.

The figure presents the results of simulations of CT with $Z=4$ and $L=10, 11, 12,\ldots, 16$.
Its pattern is similar to Fig. \ref{fig:CT_2}. In Fig. \ref{fig:CT_Z4}$\left(a\right)$ the graph presents the probability of null avalanche $P\left(S_0\right)$, as predicted by the theory and as was obtained from simulations, versus $\frac{3/2}{L}$. The scale marks of x-axis that are $\frac{3/2}{16}, \frac{3/2}{15}, \frac{3/2}{14}\ldots, \frac{3/2}{10}$ refer to CT with $L=16,15,14\ldots, 10$, respectively. For each value of $\frac{3/2}{L}$, at the top of the vertical dashed line belongs to it, a red dot is marked that represents the value of $1-\frac{3/2}{L}$ which is the probability of null avalanche $P\left(S_0\right)$ as predicted by the theory, and a blue circle is marked that represents the value of $P\left(S_0\right)$ was obtained from simulations and which was calculated to be the average of the number of null avalanches in $1000$ realizations were simulated divided by the number of attack stages $\frac{2}{3}N$. It is shown that as $L$ increases from $10$ to $16$, i.e $\frac{3/2}{L}$ decreases from $\frac{3/2}{10}$ to $\frac{3/2}{16}$, the blue circles tend to coincide with the red dots. For $L=15$ and $16$, there is indeed a good match between the blue circles and the red dots. That means that as $L$ increases, $P\left(S_0\right)$ as was obtained from simulations tends to the value predicted by our theory that is $1-\frac{3/2}{L}$.
In Fig. \ref{fig:CT_Z4}$\left(b\right)$ the graph presents the probability of complete avalanche $P\left(S_C\right)$, as predicted by the theory and as was obtained by simulations, versus $\frac{3/2}{L}$. For each value of $\frac{3/2}{L}$, at the top of the vertical dashed line belongs to it a red dot is marked that represents the value of $\frac{3/2}{L}$ which is the probability of complete avalanche $P\left(S_C\right)$ as predicted by the theory, and a blue circle is marked that represents the value of $P\left(S_C\right)$ obtained from simulations and which was calculated again to be the average of the number of complete avalanches in $1000$ realizations were simulated divided by the attack stages $\frac{2}{3}N$. It is shown that as $L$ increases from $10$ to $16$, the blue circles become close to the red dots. The inset of Fig. \ref{fig:CT_Z4}$\left(b\right)$ illustrates it again more strictly as follows -- for each of the seven CT's were simulated, the ratio of $P\left(S_C\right)$ was obtained from simulations and $\frac{3/2}{L}$ was calculated. Then the set of the $L$ values, was divided to three subsets -- $10-11, 12-14$ and $15-16$, and for each of the three subsets the mean of the ratio between $P\left(S_C\right)$ and $\frac{3/2}{L}$ was calculated, considering the members of the subset. Then a bar graph was sketched, that is the inset of Fig. \ref{fig:CT_Z4}$\left(b\right)$, where each of the three subsets is represented by one bar whose height is the mean of the ratio of $P\left(S_C\right)$ and $\frac{3/2}{L}$ was calculated for this subset. The value of this ratio is marked on top of the bar. Recall that according to our theory, as $L$ increases, $P\left(S_C\right)$ becomes greater and tends to $\frac{3/2}{L}$, that is $\frac{P\left(S_C\right)}{\frac{3/2}{L}}$ becomes greater and tends to $1$. In the bar graph was sketched, it is firstly shown that as $L$ increases between the subsets, the mean ratio of $\frac{P\left(S_C\right)}{\frac{3/2}{L}}$ becomes greater, i.e. from $0.822$ for the subset $10-11$ to $0.863$ for the subset $14-16$, and approaches towards the value $1$. 
Another evidence of the validity of the theory is related again to the prediction of the probability of complete avalanches of size $\frac{3^m-1}{2}$ to be $\frac{1}{L}\frac{1}{3^{m-1}}$ if $m\leq \log_3L$ and $0$ if $m>\log_3L$. We take for example the last subset where $L$ varies between $15$ to $16$, and respectively $\log_3L$ varies between $\log_315=2.46$ to $\log_316=2.52$. Accordingly, for this subset the theory predicts that the probability of complete avalanches of size $\frac{3^m-1}{2}$, is $\frac{1}{L}\cdot\frac{1}{3^{m-1}}$ if $m\leq 2$ and $0$ if $m>2$. With accordance to it, the probability of complete avalanche of any size $P\left(S_C\right)$ is predicted to be $\frac{1}{L}\left(\frac{1}{3^{1-1}}+\frac{1}{3^{2-1}}\right)=\frac{4/3}{L}$, and $\frac{P\left(S_C\right)}{\left(3/2\right)/L}$ is predicted to be $0.889$. In the inset of Fig. \ref{fig:CT_Z4}$\left(b\right)$ it is indeed shown that the simulation's result of $\frac{P\left(S_C\right)}{\left(3/2\right)/L}$ for this subset is $0.863$, which is $97\%$ of the predicted value $0.889$. Similarly are the results for the other subsets too.

\section{Discussion and summary}
In this work we applied a method of an analysis of an attack on networks from the microscopic perspective, that is considering the impact and the contribution of every node removal during an attack, to the fragmentation (immunization) of the network. That is in opposite to the traditional approaches that analyze attacks on networks from the macroscopic perspective, by characterizing the attack and its effectiveness by parameters related to the entire attack, such as the critical probability of percolation occurrence $p_c$ and the giant component's size. 
We found that for a random attack on CT with $Z$ neighbours of each node and a very large number of layers $L$, if we choose randomly one attack stage from the set of $\frac{Z-2}{Z-1}N$ attack stages, that is the number of node removals required in order to dismantle the network, the probability that the node removal in this chosen stage causes null avalanche is $1-\frac{\frac{Z-1}{Z-2}}{L}$.
Considering the assumption of a very large $L$, this probability is not much smaller than 1, and hence the conclusion that almost all the nodes that are removed during the attack causes null avalanche, that means that these nodes were not connected to the giant component even before they were attacked. This leads to the interesting conclusion that in our case almost all the node removals during an attack, have no effect and contribution to the destruction of the network. 
On the other hand we found that the probability that the node removal in the randomly chosen stage of the attack causes a complete avalanche is $\frac{\frac{Z-1}{Z-2}}{L}$. That is, the event of complete avalanche is the complement of the event of null avalanche. This leads to another interesting conclusion, that besides the very large part of null avalanches that are almost all the avalanches occur during an attack, there is also a small part of non-null avalanches that are of the type of complete avalanche only, and there is no non-complete avalanche during an attack. In addition to this, we have also derived explicit expressions for the distribution of the complete avalanche sizes during an entire attack on the network.

We hope that these findings will deepen our insights of how networks collapse, as a tool for improving processes of network immunization.

%\bibliography{my_bib}

\clearpage

\end{document}

\title{Supplementary Information for \\``Avalanche-size distribution of Cayley tree''}
\date{\today}
\maketitle

\renewcommand{\theequation}{A\arabic{equation}}
\setcounter{equation}{0}
\section{\raggedright I. The identity $\frac{1}{L-j+1}=\frac{1}{L+1}\sum_{m=0}^\infty \frac{j^m}{\left(L+1\right)^m}$} 
We begin with the following
\begin{equation}
	\frac{L+1}{L-j+1}=\frac{1}{1-\frac{j}{L+1}}.
\end{equation}
We assume a very large $L$, and applying the series $\frac{1}{1-x}=1+x+x^2+x^3+\dots$, so we get
\begin{align}
	\frac{1}{1-\frac{j}{L+1}}&=1+\frac{j}{L+1}+\left(\frac{j}{L+1}\right)^2+\left(\frac{j}{L+1}\right)^3+\dots\nonumber\\
	&=\sum_{m=0}^\infty\left(\frac{j}{L+1}\right)^m.
\end{align}
In summary we get
\begin{equation}
	\frac{L+1}{L-j+1}=\sum_{m=0}^\infty\left(\frac{j}{L+1}\right)^m,
\end{equation}
that is 
\begin{equation}
	\frac{1}{L-j+1}=\frac{1}{L+1}\sum_{m=0}^\infty \frac{j^m}{\left(L+1\right)^m},
\end{equation}
as required.

\renewcommand{\theequation}{B\arabic{equation}}
\setcounter{equation}{0}
\section{\raggedright II. The identity  $\sum_{j=1}^{L+1}\frac{\left(Z-1\right)^j-1}{j}=\left(Z-1\right)^{L+1}\sum_{m=0}^\infty\frac{1}{\left(L+1\right)^{m+1}}$$\sum_{j=0}^\infty\frac{j^m}{\left(Z-1\right)^j}$} 
We change variables from $j$ to $x$ according to the formula $j=L-x+1$. Therefore, $x=L-j+1$, and $j=1$ is equivalent to $x=L$ and $j=L+1$ is equivalent to $x=0$. Therefore we get
\begin{equation}
	\sum_{j=1}^{L+1}\frac{\left(Z-1\right)^j-1}{m}=\sum_{x=L}^{0}\frac{\left(Z-1\right)^{L+1-x}-1}{L+1-x}.
\end{equation} 
We rewrite the right-hand-side of the previous expression with the variable $j$, and perform the identity $\frac{1}{L-j+1}=\frac{1}{L+1}\sum_{m=0}^\infty \frac{j^m}{\left(L+1\right)^m}$. Thus we get
\begin{align}
	&\sum_{j=L}^{0}\frac{\left(Z-1\right)^{L-j+1}-1}{L-j+1}\nonumber\\
	=&\sum_{j=0}^{L}\frac{1}{L+1}\sum_{m=0}^{\infty}\frac{j^m}{\left(L+1\right)^m}\left(\left(Z-1\right)^{L-j+1}-1\right)\nonumber\\
	\approx&\sum_{j=0}^{L}\frac{1}{L+1}\sum_{m=0}^{\infty}\frac{j^m}{\left(L+1\right)^m}\cdot\left(Z-1\right)^{L-j+1},
\end{align}
where the last approximation is due to the assumption of $L>>1$. The last term can be rewritten as follows --
\begin{align}
	&\left(Z-1\right)^{L+1}\sum_{m=0}^\infty\frac{1}{\left(L+1\right)^{m+1}}\sum_{j=0}^L\frac{j^m}{\left(Z-1\right)^j}\nonumber\\
	=&\left(Z-1\right)^{L+1}\sum_{m=0}^\infty\frac{1}{\left(L+1\right)^{m+1}}\cdot\nonumber\\
	&\left(\sum_{j=0}^\infty\frac{j^m}{\left(Z-1\right)^j}-\sum_{j=L+1}^\infty\frac{j^m}{\left(Z-1\right)^j}\right).
\end{align}
Since $L>>1$, we approximate the term $\sum_{j=L+1}^\infty\frac{j^m}{\left(Z-1\right)^j}$ to $0$. Therefore we get
\begin{equation}
	\left(Z-1\right)^{L+1}\sum_{m=0}^\infty\frac{1}{\left(L+1\right)^{m+1}}\sum_{j=0}^\infty\frac{j^m}{\left(Z-1\right)^j},
\end{equation}
as required.

\renewcommand{\theequation}{C\arabic{equation}}
\setcounter{equation}{0}
\section{\raggedright III. The series $\sum_{j=0}^{\infty}\frac{j^0}{\left(Z-1\right)^j}=\frac{Z-1}{Z-2}$, $\sum_{j=0}^{\infty}\frac{j^1}{\left(Z-1\right)^j}=\frac{Z-1}{\left(Z-2\right)^2}$ and $\sum_{j=0}^{\infty}\frac{j^2}{\left(Z-1\right)^j}=\frac{Z\left(Z-1\right)}{\left(Z-2\right)^3}$ }

We begin with the familiar geometric series with the ratio $x$ where $x<1$, which is
\begin{equation}
    \sum_{j=0}^{\infty}x^j=\frac{1}{1-x}.
    \label{series_power_0}
\end{equation}
Substituting into the previous equation $x=\frac{1}{Z-1}$ gives the first required series
\begin{equation}
\sum_{j=0}^{\infty}\left(\frac{1}{Z-1}\right)^j=\frac{1}{1-\frac{1}{Z-1}}=\frac{Z-1}{Z-2}
\end{equation}

Differentiating both sides of Eq. $\left(\ref{series_power_0}\right)$ term by term gives
\begin{equation}
    \sum_{j=0}^{\infty}jx^{j-1}=\frac{1}{\left(1-x\right)^2}.
    \label{series_power_1}
\end{equation}
substituting $x=\frac{1}{Z-1}$ into Eq. $\left(\ref{series_power_1}\right)$ gives
\begin{equation}
    \sum_{j=0}^{\infty}j\left(\frac{1}{Z-1}\right)^{j-1}=\frac{1}{\left(1-\frac{1}{Z-1}\right)^2}.
\end{equation}
Performing some algebraic operations gives the second required series
\begin{equation}
    \sum_{j=0}^\infty\frac{j}{\left(Z-1\right)^j}=\frac{Z-1}{\left(Z-2\right)^2}
\end{equation}

The first term of the series in the LHS of Eq. $\left(\ref{series_power_1}\right)$ equals $0$, so it can be rewritten as follows
\begin{equation}
    \sum_{j=1}^{\infty}jx^{j-1}=\frac{1}{\left(1-x\right)^2}.
    \label{series_power_1_neg}
\end{equation}
Changing the variable j in Eq. $\left(\ref{series_power_1_neg}\right)$ to k with the rule of $j=k+1$ gives
\begin{equation}
    \sum_{k=0}^{\infty}\left(k+1\right)x^k=\frac{1}{\left(1-x\right)^2}.
\end{equation}
Rearranging this equation Considering that $\sum_{k=0}^{\infty}x^k=\frac{1}{1-x}$, yields
\begin{equation}
    \sum_{k=0}^\infty k^2x^{k-1}=\frac{1+x}{\left(1-x\right)^3}
    \label{pre_series_power_2}
\end{equation}
Substituting $x=\frac{1}{Z-1}$ into Eq. $\left(\ref{pre_series_power_2}\right)$, performing some algebraic operations and replacing formally again the variable $k$ to $j$, gives the third required identity
\begin{equation}
    \sum_{j=0}^\infty\frac{j^2}{\left(Z-1\right)^j}=\frac{Z\left(Z-1\right)}{\left(Z-2\right)^3}
\end{equation}

\renewcommand{\theequation}{D\arabic{equation}}
\setcounter{equation}{0}
\section{\raggedright IV. Changing base of expansion}
We have presented the equality between
\begin{align}
 	N\left(\frac{1}{L+1}+\frac{\frac{1}{Z-2}}{\left(L+1\right)^2}+\frac{\frac{Z}{\left(Z-2\right)^2}}{\left(L+1\right)^3}+O\left(\frac{1}{\left(L+1\right)^4}\right)\right),
\end{align}
and 
\begin{equation}
	N\left(\frac{1}{L}+\frac{\frac{3-Z}{Z-2}}{L^2}+\frac{\frac{Z^2-5Z+8}{2\left(Z-2\right)^2}}{L^3}+O\left(\frac{1}{L^4}\right)\right).
	\label{Eq_binomial_sum_5_gen}
\end{equation}

The change of the base of the expansion is implemented as folows -- we begin with the followings
\begin{equation}
	\frac{1}{n-1}=\frac{1}{n}\cdot\frac{1}{1-\frac{1}{n}}.
\end{equation} 
We apply the series $\frac{1}{1-x}=1+x+x^2+x^3+\dots$ . Therefore we get
\begin{equation}
	\frac{1}{n-1}=\frac{1}{n}\cdot\left(1+\frac{1}{n}+\frac{1}{n^2}+\frac{1}{n^3}\right).
\end{equation} 
We also write
\begin{equation}
	\frac{1}{\left(n-1\right)^2}=\frac{1}{n^2}\cdot\frac{1}{\left(1-\frac{1}{n}\right)^2}.
\end{equation} 
since the followings
\begin{equation}
	\frac{1}{\left(1-x\right)^2}=\frac{d}{dx}\left(\frac{1}{1-x}\right)=1+2x+3x^2+4x^3,
\end{equation}
then
\begin{equation}
	\frac{1}{\left(n-1\right)^2}=\frac{1}{n^2}\cdot\left(1+\frac{2}{n}+\frac{3}{n^2}+\frac{4}{n^3}\right).
\end{equation}
We write
\begin{equation}
	\frac{1}{\left(n-1\right)^3}=\frac{1}{n^3}\cdot\frac{1}{\left(1-\frac{1}{n}\right)^3}.
\end{equation}
since
\begin{equation}
	\frac{1}{\left(1-x\right)^3}=\frac{1}{2}\frac{d}{dx}\left(\frac{1}{\left(1-x\right)^2}\right)=1+3x+6x^2+10x^3,
\end{equation}
therefore we get
\begin{equation}
	\frac{1}{\left(n-1\right)^3}=\frac{1}{n^3}\cdot\left(1+\frac{3}{n}+\frac{6}{n^2}+\frac{10}{n^3}\right).
\end{equation} 
We present the equation
\begin{align}
	&\frac{a_1}{n}+\frac{a_2}{n^2}+\frac{a_3}{n^3}+\dots=\frac{b_1}{n-1}+\frac{b_2}{\left(n-1\right)^2}+\frac{b_3}{\left(n-1\right)^3}+\dots\nonumber\\
	=&\frac{b_1}{n}\cdot\left(1+\frac{1}{n}+\frac{1}{n^2}+\dots\right)+\frac{b_2}{n^2}\cdot\left(1+\frac{2}{n}+\frac{3}{n^2}+\dots\right)\nonumber\\
	&+\frac{b_3}{n^3}\cdot\left(2+\frac{3}{n}+\frac{6}{n^2}+\dots\right)+\dots=\nonumber\\
	=&\frac{b_1}{n}+\frac{b_1+b_2}{n^2}+\frac{b_1+2b_2+2b_3}{n^3}+\dots.
\end{align}
By a comparison of coefficients we get
\begin{align}
	&b_1=a_1\nonumber\\
	&b_1+b_2=a_2\quad\rightarrow\quad b_2=a_2-a_1\nonumber\\
	&b_1+2b_2+2b_3=a_3\quad\rightarrow\quad b_3=\frac{1}{2}\left(a_3-2a_2+a_1\right).
	\label{transformations_a_b}
\end{align}
In our case, $a_1=1$, $a_2=\frac{1}{Z-2}$ and $a_3=\frac{Z}{\left(Z-2\right)^2}$. Calculating the $b$'s coefficients according to the transformations in Eq. $\left(\ref{transformations_a_b}\right)$, gives that $b_1=1$, $b_2=\frac{3-Z}{Z-2}$ and $b_3=\frac{Z^2-5Z+8}{2\left(Z-2\right)^2}$ as required.

\bibliography{My_bib}

\clearpage